\documentclass[aps, prb, preprint, preprintnumbers, groupedaddress]{revtex4-1}

\usepackage{graphicx}% Include figure files
\usepackage{rotating}% rotating figure files
\usepackage{dcolumn}% Align table columns on decimal point
\usepackage{amsmath}
\usepackage[dvipsnames]{xcolor}

\begin{document}

%\preprint{}

\title{Nonlinear phononics in Bi$_2$Te$_3$ nanoscale thin films: A theoretical approach}

\author{A. Levchuk}
\altaffiliation[Present address: ]{Laboratoire des Solides Irradi\'es, UMR 7642 CEA-CNRS, Ecole Polytechnique, France, EU.}
\author{R. Busselez}
\author{G. Vaudel}
\author{P. Ruello}
\author{V. Juv\'e}
\author{B. Arnaud}
\email[e-mail: ]{brice.arnaud@univ-lemans.fr}
\affiliation{Institut des Mol\'ecules et Mat\'eriaux du Mans, UMR CNRS 6283, Le Mans universit\'e, 72085 Le Mans, France, EU}


%\author{}
%\email[]{brice.arnaud@univ-rennes1.fr}
%\thanks{}
%\affiliation{}


\date{\today}

\begin{abstract}
Density Functional Theory (DFT) calculations not only allow to predict
the vibrational and optical properties of solids but also to understand 
and disentangle the mechanisms playing a key
role in the generation of coherent optical phonons.
Recent experiments performed on a Bi$_2$Te$_3$ nanoscale thin film
have shown that a THz pulse launches at least a coherent $A_{1g}^1$ phonon 
as the transient transmittance measured using an isotropic detection 
scheme displays oscillations with a frequency matching the frequency of 
the $A_{1g}^1$ mode measured in Raman experiments.
Such an observation can be explained by invoking either 
a sum frequency process or cubic/quartic phonon-phonon couplings as considered
for Bi$_2$Se$_3$, a parent compound of Bi$_2$Te$_3$. 
By resorting to group theory and
calculating energy surfaces from first-principles,
the main phonon-phonon couplings can be identified. 
Furthermore, a minimal model can be built to compute the dynamics
of the Raman active modes coupled to the infrared active mode driven 
by the experimental THz pulse.
%to explain
%pump-probe experiments. 
%Our model is fully validated 
Our model firmly establishes that cubic phonon-phonon interactions are relevant 
%explain
%the explain the experimental observation as
as the agreement between the computed and 
experimental transmittance is noteworthy.

%This model is validated
%by simulating the detection process as the modification 
%of the dielectric function
%arising from the generation of optical phonons can easily
%be evaluated at the random phase approximation (RPA) level.
%Therefore, ab-initio calculations when combined with models
%are invaluable tools to shed lights on the complex mechanisms
%at the heart of ultrafast physics.
\end{abstract}

%\pacs{63.20.dk, 65.40.De}
%\keywords{}

\maketitle
\section{INTRODUCTION}
Advances in THz physics\cite{kampfrath_2013}, 
through the generation\cite{Reimann_2007, Kitaeva_2008, Hebling_2008, Hirori_2001} 
and detection\cite{Wu_1995, Brunner_2014, Johnson2_2014} of
intense THz pulses, have revolutionized the field of condensed matter physics,
where the concept of quasiparticles like excitons, magnons, and phonons, is
essential. Indeed, low energy excitations can be targeted to 
induce non-equilibrium quantum states with novel properties that challenge 
our understanding of solid state physics. Phonons are quasiparticles with
a finite lifetime essentially arising from phonon-phonon interactions at
high temperature. These interactions, which provide a natural explanation for the 
finite thermal conductivity of solids, are leveraged in the blooming field
of nonlinear phononics\cite{forst_2011, Subedi_2014, Disa_2021}. In a few words, a THz pulse 
%with a peak field strength up to a few MeV.cm$^{-1}$ 
drives resonantly or nonresonantly one or more 
infrared (IR) active modes which in turn
are coupled to other symmetry allowed modes. 
To cite few examples\cite{Nicoletti_2016}, nonlinear phonon-phonon mixing has been exploited to 
induce insulator-to-metal transition in a manganite\cite{Rini_2007}, 
to reverse the polarization
in a ferroelectric\cite{Mankowsky_2017} or even to induce ferrimagnetism 
in an antiferromagnet by 
mimicking the effect of an externally applied strain\cite{Disa_2020}.
%Thus, the dynamics of atoms is steered on the 
%femtosecond time scale and transient states, with no equilibrium counterpart, might
%be induced.

% 1 Our recent THz-pump optical-probe experiments
% 1 performed on a $\sim 12$ nm thick Bi$_2$Te$_3$ nanofilm 
% 1 deposited on a mica substrate\cite{Levchuk_2020} 
% 1 have shown that the $A_{1g}^1$ phonon with a frequency $\sim 1.86$ THz
% 1, also seen in in Raman experiments\cite{Richter_1977, Wang_2013}, 
% 1 is coherently driven by a THz pulse with a frequency centered on $\sim 0.6$ THz.

Our recent THz-pump optical-probe experiments
performed on a $\sim 12$ nm thick Bi$_2$Te$_3$ nanofilm
deposited on a mica substrate\cite{Levchuk_2020}
have shown that the measured transient transmittance displays a 
fast oscillatory component with a frequency $\sim 1.86$ THz
which can be ascribed to the coherently driven $A_{1g}^1$ optical phonon 
seen in Raman experiments\cite{Richter_1977, Wang_2013}.
%seen in Raman experiments\cite{Richter_1977, Wang_2013}
%In this Letter, we provide a theoretical interpretation of the 
%pump-probe experiments performed on a n-doped 
%$\sim 12$ nm thick
%Bi$_2$Te$_3$ nanofilm deposited on a mica substrate\cite{weis_2015, Levchuk_2020, SM}. 
%deposited on a mica substrate\cite{weis_2015, Levchuk_2020, SM}. 
%As shown in Fig. \ref{Delta-T-fig}(a), the free electrons are driven by the THz
%pump pulse shown in the inset and give rise to a steep increase of the transmittance
%followed by a decay arising from electron-phonon scattering 
%events\cite{Allen_1987, Arnaud_2013} on a time scale $\sim 4$ ps. 
%The fast oscillatory component with a frequency $\sim 1.86$ THz, that is surimposed
%on the decay, can be ascribed to the coherently driven $A_{1g}^1$ optical phonon 
%seen in Raman experiments\cite{Richter_1977, Wang_2013}.
% while the low frequency 
%oscillations are related to the launching of longitudinal acoustic waves. 
It's worth remarking that our isotropic detection scheme is only sensitive 
to the symmetry preserving modes, namely the $A_{1g}$ modes. Interestingly,
Melnikov {\it{et al}}\cite{Melnikov_2018} performed similar experiments 
on a Bi$_2$Se$_3$ nanofilm
and observed the $A_{1g}^1$ mode by measuring the transient transmittance 
as well as both the $E_u^1$, $E_g^1$ and $E_g^2$ modes by measuring 
the transient polarization rotation. The generation of the $E_g^2$ 
($A_{1g}^1$ and $E_g^1$) mode was tentatively explained by invoking a
third (fourth) order coupling with the $E_u^1$ mode driven by the THz pulse.
Thus, the question arises to know whether quartic interactions, as speculated
for Bi$_2$Se$_3$ which is a parent compound of Bi$_2$Te$_3$, might explain 
the generation of the $A_{1g}^1$ phonon mode seen in our experiments. 

{\textit{Ab initio}} calculations have already shown their strength in disentangling 
the many different processes occuring in the field 
of ultrafast physics\cite{Giret_2011, Arnaud_2013, Mahony_2019, Sangalli_2015, Sjakste_2021, Tong_2021}. 
Furthermore, they offer not only the possibility to identify 
relevant mechanisms by evaluating the phonon-phonon coupling terms\cite{Subedi_2014, Kozina_2019} 
but also provide the less explored opportunity to simulate the detection process,  
allowing a direct and quantitative comparison with the experimental results. 
%Our aim is to provide a seamless theoretical interpretation of our 
%pump-probe experiments\cite{Levchuk_2020}.

The paper is organized as follows. In section \ref{expt-part},
we briefly describe the experimental setup.
%. In section \ref{pulse-part},
We then show how the temporal profile of the THz pulse 
%can be extracted 
is extracted
from electro-optic measurements 
and accurately fitted with an analytical function. Particular emphasis is placed on
the precise determination of the electric field amplitude,
%with particular emphasis on precisely determining the electric field 
%amplitude, 
a parameter of the uttermost importance for our simulations.
In section \ref{Expt_fitting}, we explain how the oscillatory part of the transient transmittance
arising from the launching of the coherent A$_{1g}^1$ phonon is extracted
from the raw measured transient transmittance. In section \ref{Technical_part}, 
we describe the crystallographic structure of Bi$_2$Te$_3$ and give an account 
of the technicalities used to perform our first-principles calculations.
In section \ref{Sample_part}, we describe how the experimentally studied heterostructure,
composed of an oxidized Bi$_2$Te$_3$ thin film deposited on a mica substrate, can 
be characterized in terms of layer thicknesses. For this purpose, the experimental measurement
of the heterostructure's wavelength-dependent transmittance is analysed using a transfer 
matrix method, where knowledge of each layer's thickness and optical constants allows theoretical
calculation of the transmittance.
In section \ref{energy_surface}, we apply group theory to analyze the dominant
zone-center phonon-phonon couplings, as derived from 
energy surface calculations. In section \ref{Eq_phonon_dynamics_part}, we derive the equations
of motion governing the phonon modes involved in our THz pump-optical probe experiments.
In section \ref{Dynamics_modes_part}, we present the dynamics of the zone-center modes
following the arrival of the THz pulse and demonstrate 
how the carrier-envelope phase can be manipulated to influence the long-time evolution
of these modes. In section \ref{detection_part}, we describe the treatment of the detection
process and provide a direct comparison between the computed and experimentally measured transient transmittance at 400 nm.
In Section \ref{Raman_forces_part}, 
we compute the forces related to two-photon absorption processes, 
also referred to as Raman sum-frequency processes, 
and show that these forces can be neglected  in favor 
of those arising from lattice anharmonicity. Thereby, we justify the
omission of these forces in the simulations presented 
in Section \ref{Dynamics_modes_part}.
In Section \ref{analytical_part}, we derive an analytical formula describing 
the long-time behavior of the dynamics of the $E_u^1$ modes 
and validate this formula in Section \ref{hypothetical_pulse_part}, 
where we demonstrate that it is possible to transiently lower
the symmetry of Bi$_2$Te$_3$ by exploiting an hypothetical 
THz pulse resonant with the E$_u^1$ modes. In Section \ref{conclusion_part},
we summarize our key findings.

\section{Experimental setup and THz pulse characterization}\label{expt-part}
The experiments employ an ultrafast THz pump–optical probe setup 
in the transmission geometry.
%illustrated in Fig.\ref{fig1}. 
Both the pump and probe pulses originate from an amplified Ti:sapphire laser 
system that delivers 150-fs pulses centered at 800 nm, 
with a pulse energy of 3 mJ and a repetition rate of 1 kHz. 
The majority of the pulse energy is used to generate intense THz radiation, 
while the residual 800 nm light is frequency-doubled in a BBO crystal 
and used as the probe. THz pulses are produced via optical rectification 
in a LiNbO$_3$ crystal\cite{Hirori_2001,Levchuk_2020}, 
and their temporal profile is characterized through electro-optic sampling\cite{Brunner_2014, Johnson2_2014} using a 200-$\mu$m-thick GaP crystal.

As our simulations aim at quantitatively reproduce the experimental 
results, a special attention was devoted to the characterization of the THz 
electric field shown in Fig. \ref{fig2}(a) and whose
spatio-temporal profile is well approximated by:
\begin{equation}
	E(x, y, t)= E_0(t) g(x,y),
\end{equation}
where
\begin{equation}\label{E0_vs_t}
	E_0(t)=E_0 \sin\left[\omega_0(t-t_0) + \Phi_0 \right] 
\exp\left[-\frac{(t-t_0)^2}{\sigma^2}\right]
\end{equation}
is the time-domain waveform of the THz electric field.
%retrieved from
%the electro-optic measurements\cite{Brunner_2014, Johnson2_2014}.
\begin{figure}[!hbt]
\begin{center}
\vskip0.5truecm
\includegraphics[angle=0, scale=0.3]
{./FIG-1-PRB.eps}
\end{center}
%\vskip -0.75truecm
\caption{
\label{fig2}
(a) 
The THz electric field (open circles) 
measured using an electro-optic 
method\cite{Brunner_2014, Johnson2_2014}, 
is compared to a fit based on the
analytical function given
by Eq. \ref{E0_vs_t} (solid black line).
(b) Fourier transforms of both the experimental (open circles) and analytical 
(solid black line) 
THz waveform  together with the computed zone
center frequencies at the LDA level shown as vertical arrows\cite{busselez_2023}.
}
\end{figure}
Fig. \ref{fig2}(a) shows that the experimental THz waveform is well fitted
with $\omega_0/2\pi=0.64$ THz, $\sigma=0.79$ ps and $\Phi_0=0.117 \pi$.
Note that $t_0$ is just a reference time which is set to zero in our simulations.
For the sake of completeness, the Fourier transforms (FT) of 
both the experimental and analytical THz waveform are displayed in Fig. \ref{fig2}(b).
As expected, the FT exhibit a peak centered on $\sim 0.64$ THz.
The spatial profile of the THz electric field, which is assumed to be Gaussian, 
is given by the following expression:
\begin{equation}
g(x,y)=\exp\left[-\frac{(x-x_0)^2}{w_x^2 } \right]
\exp\left[-\frac{(y-y_0)^2}{w_y^2 } \right],
\end{equation}
where
$w_x=0.883\pm 0.048$ mm and $w_y=0.572\pm 0.048$ mm are both obtained
by imaging the THz beam with a pyroelectric infrared camera 
which consists of a $320\times 240$
pixel imaging array with a pixel size of $48.5~\mu$m. The energy per pulse,
denoted as $W$, which is given by $W=P/\nu$ where $P=1.2$ mW is the power 
measured by a pyroelectric detector and $\nu=1$ kHz is the repetition
rate, can also be written as:
\begin{equation}
W=\frac{1}{\mu_0 c} \int dt \iint dx dy~ E(x,y, t)^2=
\frac{E_0^2}{\mu_0 c} \underbrace{\iint dx dy~ g(x, y)^2}_{S=w_x w_y/2} 
\underbrace{\int dt f(t)^2}_K.
\end{equation}
Hence, we get:
\begin{equation}\label{def_E0}
E_0=\sqrt{\frac{ \eta_0 W}{S K}},
\end{equation}
where $\eta_0=\mu_0 c$ is the free space impedance ($\eta_0=377 ~\Omega$). 
Using Eq. \ref{def_E0} and assuming that $\Delta W/W=10$ \%, we
obtain $E_0=340 \pm 40$ kV.cm$^{-1}$ as 
$\Delta E_0/E_0=[\Delta W/W + \Delta S/S]/2\sim 0.12$. The simulations
discussed below are performed for $E_0=340$ kV.cm$^{-1}$.

\section{Experimental data fitting}\label{Expt_fitting}

The measured transient relative transmittance $\Delta T(t)/T$ of the sample
shown in Fig. \ref{fit_fig} is nicely fitted by the following expression:
%\begin{eqnarray}\label{fit_eq}
%\Delta T(t)/T_0 = \frac{1}{2}[\textrm{erf}(\Delta t/s) + 1]
%\left\{a_1\exp(-\Delta t/\tau_{ep}) +  
%a_2\exp(-\Delta t/\tau_{hd})  \right. \nonumber \\
%\left.+\:a_3\sin(\omega_{ac}\Delta t +\phi_{ac} )
%\exp(-\Delta t/\tau_{ac}) + 
%a_4\sin(\omega_{opt}\Delta t +\phi_{opt} ) 
%\exp(-\Delta t/\tau_{opt}) \right\},
%\end{eqnarray}
\begin{eqnarray}\label{fit_eq}
\Delta T(t)/T = \frac{1}{2}[\textrm{erf}(\Delta t/s) + 1]
\left\{a_1\exp(-\Delta t/\tau_{ep}) +  
a_2\exp(-\Delta t/\tau_{hd})  \right. \nonumber \\
\left.+\:a_3\sin(2\pi\nu_{ac}\Delta t +\phi_{ac} )
\exp(-\Delta t/\tau_{ac}) + 
a_4\sin(2\pi\nu_{opt}\Delta t +\phi_{opt} ) 
\exp(-\Delta t/\tau_{opt}) \right\},
\end{eqnarray}
where $\Delta t = t-t_0$ ($t_0$ is a reference time) and erf is the error function.
\begin{figure}[!htbp]
\vskip1.0truecm
\includegraphics[angle=0, scale=0.4]
{./FIG-2-PRB.eps}
%    \centering 
%    \includegraphics[width=0.98\textwidth]{Figure_Fit_SM.eps}
	\caption{(a) Transient experimental relative transmittance $\Delta T/T$ 
(black line) measured as a function of the time delay $t$ between the optical pulse
and the THz pulse compared to the fitting curve (red dashed line) obtained from Eq. \ref{fit_eq} with
the parameters reported in Table \ref{fit_expt_tab}. The difference between the experimental
and the fitting curve corresponds to the blue dotted curve.
(b) Experimental curve (black line) compared to the background curve 
(red dashed line) 
obtained by setting $a_4=0$ in Eq. \ref{fit_eq} while keeping the values of the other
parameters found in Table \ref{fit_expt_tab}. The difference between the experimental
and the background curve provides the high frequency part of the signal (blue dotted curve) 
related to the coherent optical phonon oscillations. The signal to noise ratio 
of the experimental relative transmittance is $\sim 200$.}
\label{fit_fig}
\end{figure}
All the parameters entering Eq. \ref{fit_eq} are determined by a least squares
fit method. The rising time of the 
signal is governed by the parameter $s=0.21$ ps 
as the function $t \mapsto (\textrm{erf}(t/s)+1)/2$ looks like a smoothed heaviside
function. All the other parameters are gathered in table \ref{fit_expt_tab}. 
%\begin{table}[!hbt]
%\caption{ Parameters used in Eq. \ref{fit_eq} to best fit the transient experimental
%relative transmittivity shown in Fig. \ref{fit_fig}.
%} 
%\label{fit_expt_tab}
%\begin{tabular}{cccc}
%\hline
%   \multicolumn{4}{c}{Electronic contribution} \\
% $a_1$ & $\tau_{ep} (ps) $ & $a_2$ & $\tau_{hd}$ (ps) \\ 
%0.343  & 4.0             & 0.068   & 128.82    \\
%\hline
%   \multicolumn{4}{c}{Acoustic phonon contribution} \\
%$a_3$ & $\omega_{ac}/2\pi$ (Ghz) & $\phi_{ac}$ & $\tau_{ac} (ps)$  \\
%0.069 & 76.65                    & -0.127      & 13.18             \\ 
%\hline
%   \multicolumn{4}{c}{Optical phonon contribution} \\
%$a_4$ & $\omega_{opt}/2\pi$ (Thz) & $\phi_{opt}$ & $\tau_{opt} (ps)$ \\
%0.08  & 1.86                      & -0.960    & 3.25 \\
%\hline
%\end{tabular}
%\end{table}
\begin{table}[!hbt]
\caption{ Parameters used in Eq. \ref{fit_eq} to best fit the transient experimental
relative transmittance shown in Fig. \ref{fit_fig}.
} 
\label{fit_expt_tab}
\begin{tabular}{cccccccccccc}
\hline
   \multicolumn{4}{c}{Electronic contribution} & \multicolumn{4}{c}{Acoustic phonon contribution} & \multicolumn{4}{c}{Optical phonon contribution} \\
 $a_1$ & $\tau_{ep} (ps) $ & $a_2$ & $\tau_{hd}$ (ps) & $a_3$ & $\nu_{ac}$ (Ghz) & $\phi_{ac}$ & $\tau_{ac} (ps)$ & $a_4$ & $\nu_{opt}$ (Thz) & $\phi_{opt}$ & $\tau_{opt} (ps)$\\ 
0.343  & 4.0             & 0.068   & 128.82    & 0.069 & 76.65                    & -0.127      & 13.18 & 0.08  & 1.86                      & -0.960    & 3.25 \\
\hline
\end{tabular}
\end{table}
From a physical point of view, the free electrons of the Bi$_2$Te$_3$ nanoscale
thin film are driven by the THz pulse and give rise
to a steep increase of the transmittance followed by a decay arising from 
electron-phonon scattering events\cite{Allen_1987, Arnaud_2013} 
on a time scale $\sim 4$ ps and by a slower
decay arising from heat diffusion on a time scale $\tau_{hd}\sim 129$ ps. 
Furthermore, the transmittance displays high frequency damped oscillations with
a frequency $\nu_{opt} \sim 1.86$ THz and a damping time 
$\tau_{opt}\sim 3.25$ ps respectively matching 
the zone center A$_{1g}$ phonon frequency
$\nu_{A_{1g}} \sim 1.86$ THz 
(see Table \ref{effective_charge_tab}) 
and lifetime $\tau_{A_{1g}} \sim 3.54$ ps 
extracted from Raman measurements\cite{Wang_2013, vilaplana_2011}.
Our primary theoretical objective is to quantitatively model
the oscillatory transient transmittance displayed as a blue dotted line
in Fig. \ref{fit_fig}(b) through a multiscale methodology bridging
{\it{ab initio}} calculations and models. This task requires 
a comprehensive description of both the coherent phonon generation
mechanism and its detection scheme.

\section{Crystallographic structure and technical details for {\it{ab initio}} calculations}\label{Technical_part}
Bi$_2$Te$_3$ crystallizes in a rhombohedral structure, also called A7 structure,
with five atoms per unit cell\cite{francombe_1958, jenkins_1972, busselez_2023}.
%The vectors ${\bf{a}}_1$, ${\bf{a}}_2$ and
%${\bf{a}}_3$ spanning the unit cell
%at 300 K\cite{francombe_1958} have a length $a=10.473$ \AA~
%and the angle between any pair of vector is $\alpha=24.159^\circ$.
The three Te atoms can be classified into two inequivalent types.
Two of them, labelled as Te$_1$, are located at
$\pm \nu {\bf{a}}_\parallel$ while the last
Te atom, labelled as Te$_2$, is set at the origin.
The two Bi atoms are equivalent and located
at $\pm \mu {\bf{a}}_\parallel$. Here,
$\nu$ and $\mu$ are dimensionless parameters and
${\bf{a}}_\parallel={\bf{a}}_1+{\bf{a}}_2+{\bf{a}}_3$
is parallel to the trigonal axis ($C_3$ axis), where ${\bf{a}}_1$, 
${\bf{a}}_2$
and ${\bf{a}}_3$ span the rhombohedral unit cell\cite{busselez_2023}.
\begin{figure}[!htbp]
\vskip1.0truecm
\includegraphics[angle=0, scale=0.35]
{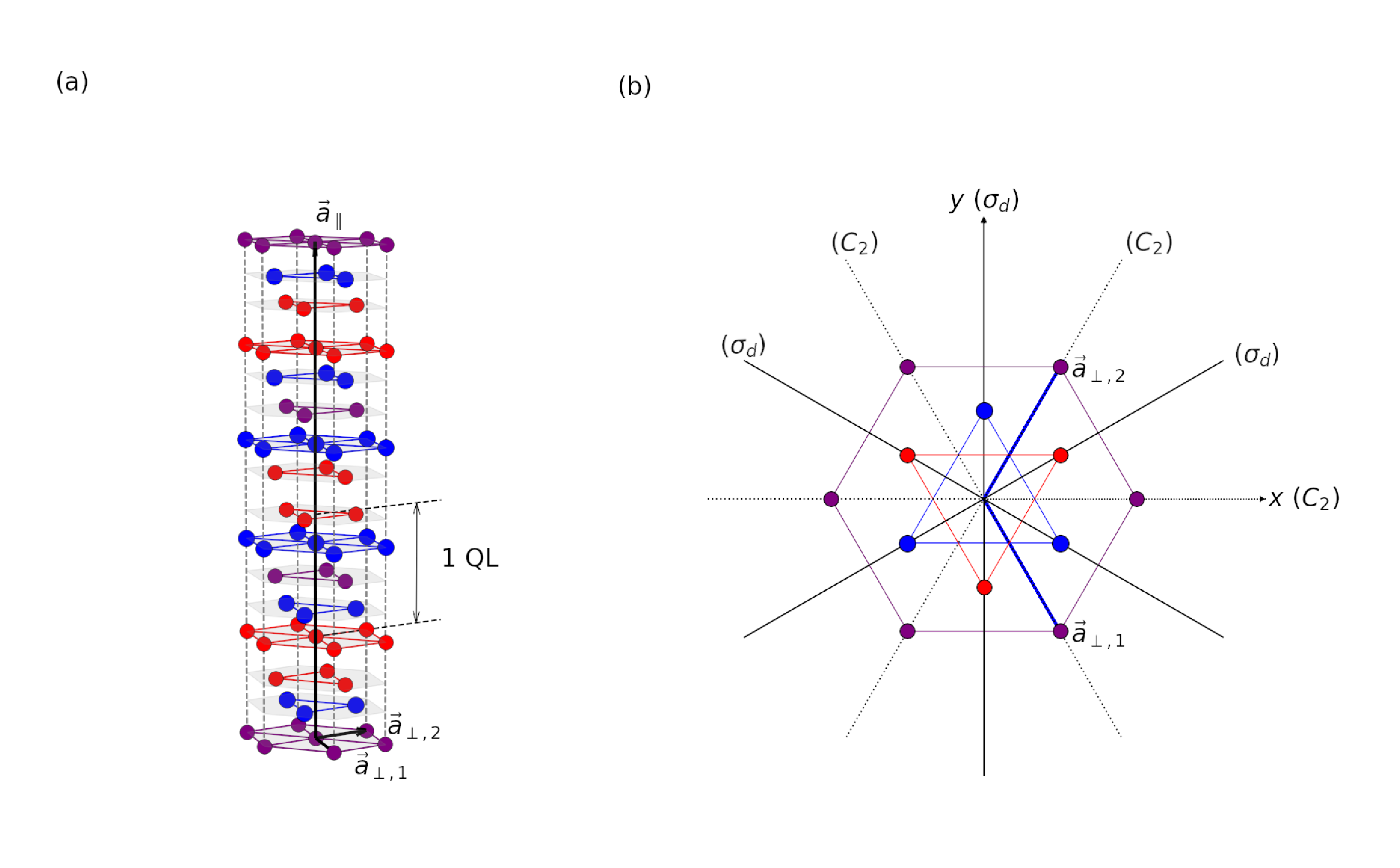}
\caption{(a) Hexagonal structure of Bi$_2$Te$_3$ spanned by
the lattice vectors ${\bf{a}}_{1, \perp}$, ${\bf{a}}_{2, \perp}$
and ${\bf{a}}_{\parallel}$. The Te$_1$ atoms, Te$_2$ atoms
and Bi atoms are respectively colored in red, purple and blue.
The double-headed arrow delineates one quintuple layer (QL) 
with a thickness of approximately $0.76$ nm.
(b) Top view of the hexagonal structure with the three $\sigma_d$ planes
and $C_2$ axis respectively represented by straight and dotted lines.}
\label{Structure_fig}
\end{figure}
Alternatively, the structure can be viewed as a hexagonal structure
depicted in Fig. \ref{Structure_fig}(a) and made of three quintuple layers
Te$_1$-Bi-Te$_2$-Bi-Te$_1$ with a a thickness of approximately $0.76$ nm 
separated by so called Van der Waals gaps with thickness of approximately 
$0.26$ nm (distance between two successive planes made of Te$_1$ atoms).

All the ground state calculations are performed for the experimental
lattice parameters at room temperature\cite{francombe_1958}
within the framework of the local
density approximation (LDA)
%for the exchange-correlation functional to density functional theory (DFT) 
as implemented in the ABINIT code\cite{gonze_2009, gonze_2016}.
Relativistic separable dual-space Gaussian
pseudo-potentials\cite{hartwigsen_1998} are used with Bi
($6s^2 6 p^3$) and Te ($5s^2 5 p^4$) levels treated as valence states.
Spin-orbit coupling was included
and an energy cut-off of 40 Hartree in the planewave expansion of wavefunctions
as well as a $16\times 16\times 16$ kpoint grid for the
Brillouin zone integration were used.

%The optical properties including local field (LF) effects at the Random Phase Approximation (RPA)
The optical properties at the Random Phase Approximation (RPA)
level\cite{arnaud_2001} are computed with the YAMBO code\cite{yambo_2009, yambo_2019}
which requires the ground state electronic structure computed with
the ABINIT code\cite{gonze_2009, gonze_2016}.  From a practical point of view,
we used a $64\times 64 \times 64$ kpoints grid (22913 irreducible kpoints) and
included 28 valence bands and 36 conduction bands in our calculations to converge the dielectric function.
%without local field (LF) effects. 
It's worth pointing out that local field effects can be discarded as they only
slightly affect the in-plane optical properties\cite{busselez_2023}.
%Regarding the inclusion of LF effects, 302 reciprocal
%lattice vectors are included in the calculation of the body matrix to ensure
%the convergence of the macroscopic dielectric function\cite{busselez_2023}.

\section{Characteristics of the sample}\label{Sample_part}
Our measurements are carried out on a Bi$_2$Te$_3$ nanoscale thin film 
deposited on 
a mica substrate\cite{Levchuk_2020}. The thin film is covered by an oxidized 
layer whose exact nature and thickness remain unknown. 
However, X-ray photoelectron spectroscopy experiments\cite{weis_2015}
have shown the presence of Bi-O and Te-O bonds that are the hallmarks of 
an oxidized layer displaying the properties of a glass\cite{bando_2000} 
which might share some properties with bulk Bi$_2$O$_3$ and bulk TeO$_2$.
As shown in Fig. \ref{fit_fig}, the transmittance also displays 
low frequency oscillations ($\nu_{ac}=76.65$ GHz)
related to the launching of coherent longitudinal acoustic waves that are totally reflected at the
 Air/Oxidized layer interface and partially reflected 
at the Bi$_2$Te$_3$/mica interface\cite{weis_2015}. The analysis of the
acoustic waves is a first step towards a correct description of the heterostructure.

%The analysis of the datas 
%is more difficult because of the existence of an oxidized layer covering the
%Bi$_2$Te$_3$ nanofilm. 
%However,
%X-ray photoelectron spectroscopy experiments\cite{bando_2000}
%have shown the presence of up to two oxidized quintuple layers (QL) 
%with Bi-O and Te-O bonds after 6000 hours of exposure to air. 

As we do
not know the sound velocity, denoted as $\overline{V}$, of the longitudinal
waves inside the oxidized layer, we assume that $\overline{V}=\alpha V$, 
where $\alpha$ is a dimensionless coefficient and $V=\sqrt{C_{33}/\rho}$ 
is the longitudinal sound velocity along the trigonal axis of Bi$_2$Te$_3$. 
As the room temperature elastic contant\cite{jenkins_1972} 
and density\cite{francombe_1958} are 
respectively $C_{33}=47.68$ GPa and $\rho=7863$ kg.m$^{-3}$, 
we obtain $V=2462$ m.s$^{-1}$. Hence, we can write:
\begin{equation}
\frac{1}{\nu_{ac}} = \frac{2 d_0}{V} \left[ \frac{m}{\alpha} + 3 n\right],
\end{equation}  
where $d_0~\sim~1.01$ nm is the thickness of one block 
(one QL + one Van der Waals gap), 
$n$ is the number of stacked unit cells
(As depicted in Fig. \ref{Structure_fig}(a), 
one unit cell includes three blocks) 
and $m$ is the number of oxidized blocks ($m=1, \cdots, 4$). 
\begin{figure}[!htbp]
\vskip0.5truecm
\includegraphics[angle=0, scale=0.3]
{./FIG-4-PRB.eps}
\caption{$f_m$ (See Eq. \ref{Eq_fm}) as a function of $\alpha=\overline{V}/V$,
where $\overline{V}$ ($V$) is the longitudinal sound velocity inside the 
oxidized layer (Bi$_2$Te$_3$ film). Here $m=1, \cdots, 4$ denotes the number of
oxidized blocks. The horizontal lines show the only allowed integer value of $n$
(number of unit cells) and the vertical lines show that the only realistic value
of $m$ compatible with the values of $n$ are one and four.  }
\label{Layers_fig}
\end{figure}
Thus, the Bi$_2$Te$_3$ film thickness is given by $d=3n d_0$
%while the thickness of the oxidized layer is approximated by 
%$d_{ox}=m d_0$, under the simplifying assumption 
%that it remains constant throughout the oxidation process.
%while the thickness of the oxidized layer is approximately given by 
%$d_{ox}=m d_0$ . This estimate is, however, a rough approximation, 
%as the thickness of the oxidized block is expected to vary 
%during the oxidation process.
while the thickness of the oxidized layer is approximated by 
$d_{ox}=m d_0$ under the simplifying assumption that the
thickness of the oxidized block remains constant 
throughout the oxidation process.
%This estimate is, however, a rough approximation, 
%as the thickness of the oxidized block is expected to vary 
%during the oxidation process.
%Thus, the oxidized layer and Bi$_2$Te$_3$ film thicknesses are respectively 
%given by $d_{ox}=m d_0$ and $d=3n d_0$. 
The allowed values of $n$ are given by the 
integer values of the function $f_m(\alpha)$ defined as:
\begin{equation}\label{Eq_fm}
f_m(\alpha)=\frac{1}{3} \left[ \frac{V}{2 d_0 \nu_{ac}} -\frac{m}{\alpha}\right].
\end{equation}
Fig. \ref{Layers_fig} shows that either $n=4$ with $m=4$ ($d\sim 12.19$ nm
and $d_{ox} \sim 4.06$ nm) or $n=5$ with $m=1$ 
($d\sim 15.24$ nm and $d_{ox} \sim 1.01$ nm) are acceptable
given the fact that $\overline{V}=1.05 V$ or $\overline{V}=1.24 V$
are compatible with
the sound velocity inside the oxidized layer.
\begin{figure}[!htbp]
\vskip1.0truecm
\includegraphics[angle=0, scale=0.4]
{./FIG-5-PRB.eps}
%    \centering 
%    \includegraphics[width=0.98\textwidth]{Figure_Fit_SM.eps}
\caption{
(a) Real and imaginary parts of the optical index of Bi$_2$Te$_3$ 
computed at the RPA level\cite{arnaud_2001, yambo_2019, busselez_2023} 
for an electric field perpendicular to the trigonal axis as a function of the
wavelength $\lambda$ (in nm). (b) Experimental real part of the optical index
of Bi$_2$O$_3$\cite{dolocan_1981}, TeO$_2$\cite{uchida_1971} and mica\cite{nitsche_2004} 
as a function of $\lambda$ (in nm). (c) Computed transmittance of the heterostructure
corresponding to $n=5$ ($d\sim 15.24$ nm) and $m=1$ ($d_{ox}\sim 1.01$ nm) as a function
of $\lambda$ (in nm) for an oxide layer made either of Bi$_2$O$_3$ (purple curve) 
or TeO$_2$ (red curve) compared to the experimental transmittance (black curve) 
measured with a white lamp (black curve). (d) Same as (c) but for $n=4$ ($d\sim 12.19$ nm)
and $m=4$ ($d_{ox}\sim 4.06$ nm). The two possible heterostructures are schematically 
represented in (c) and (d). The scale for the mica layer thickness ($d_{mica}\sim 17.1 \mu$m)  
is not respected as it is much larger than the oxide and nanofilm thicknesses.
}
\label{Optical_index_fig}
\end{figure}
%In order to choose the correct structure
In order to specify the correct thicknesses,
 we measured the transmittance of
the experimental heterostructure using a white lamp. 
As shown in Fig. \ref{Optical_index_fig}(c,d), the transmittance decreases as $\lambda$
increases (from the blue to the red part of the spectral range) and displays some
oscillations related to interference effects taking place within the partially incoherent
mica layer. The mica thickness extracted from the interference patterns is
$d_{mica}\sim 17.1 \mu$m. Using a transfer matrix approach\cite{katsidis_2002} 
and the complex
refractive index of Bi$_2$Te$_3$ (see Fig. \ref{Optical_index_fig}(a)) 
evaluated at the RPA level\cite{arnaud_2001, yambo_2019, busselez_2023}
%where local field effects
%(LF) are included
as well as 
the experimental optical index of Bi$_2$O$_3$\cite{dolocan_1981}, 
TeO$_2$\cite{uchida_1971} and mica\cite{nitsche_2004} 
(See Fig. \ref{Optical_index_fig}(b)), we
compute the transmittance of both envisionned heterostructures where the mica layer
is treated as an incoherent layer. As stated before, the optical properties of the
oxidized layer are unknown. Thus, we assume that the oxidized layer is made 
either of Bi$_2$O$_3$ or of TeO$_2$. As shown in Fig. \ref{Optical_index_fig}(c) 
and Fig. \ref{Optical_index_fig}(d), the computed transmittance is nearly the
same in both cases. However, the computed transmittance for
$n=5$ ($d\sim 15.24$ nm) and $m=1$ ($d_{ox}\sim 1.01$ nm)   
displayed in Fig. \ref{Optical_index_fig}(c) is underestimated with respect 
to experiment while the computed transmittance for
$n=4$ ($d\sim 12.19$ nm) and $m=4$ ($d_{ox}\sim 4.06$ nm) 
displayed in Fig. \ref{Optical_index_fig}(d) is in fairly good agreement
with respect to experiment.
Finally, we 
conclude that the thickness of the Bi$_2$Te$_3$ film is 
$d \sim 12.19$ nm (4 unit cells) and
that the film is covered
with a 4.06 nm thick oxidized layer (4 oxidized blocks). These
results are in line with
both the ex situ x-ray reflectivity measurements and the quartz crystal microbalance
measurements\cite{weis_2015}.

\section{Energy surfaces}\label{energy_surface}
As the primitive cell of Bi$_2$Te$_3$ contains five atoms, 
there are 15 lattice dynamical modes at
${\bf q}={\bf{0}}$, three of which are acoustic modes. 
Group theory classifies the
remaining 12 optical modes into 2 A$_{1g}$ (R), 2 E$_g$ (R), 2 A$_{2u}$ (IR) 
and 2 E$_{u}$ (IR) modes, where R and IR refer to Raman and infrared 
active modes respectively.
The dynamical matrix is computed using density functional perturbation
theory\cite{gonze_1997} and diagonalized to get the zone center frequencies 
%$\omega_\lambda({\bf q}={\bf 0})$ 
shown in Fig. \ref{fig2}(b) 
as downward (upward) arrows for IR (R) active modes. 
%The overall agreement between
%theory\cite{busselez_2023} and experiment\cite{richter_1977} is reasonable. 
%\textcolor{blue}{The only modes that can be coupled to a THz pulse polarized perpendicularly
%to the trigonal axis ($\bf{E} \perp {\bf{a}}_\parallel$) 
%are the E$_u^1$ and E$_u^2$
%modes. However, Fig. \ref{fig2}(b) shows that the spectral amplitude 
%of the THz pulse is completely negligible for the E$_u^2$ mode. Therefore,
%the E$_u^1$ mode is the only IR active mode that can be driven non-resonantly
%by the THz pump pulse.}
Only the E$_u^1$ and E$_u^2$ modes can couple to 
a THz pulse polarized perpendicular 
to the trigonal axis ($\bf{E} \perp {\bf{a}}_\parallel$).
However, as shown in Fig. \ref{fig2}(b), the spectral amplitude 
of the THz pulse is entirely negligible for the E$_u^2$ mode. Consequently,
the E$_u^1$ mode remains the sole IR-active mode that 
can be driven 
%non-resonantly
by the THz pump pulse.

We now focus on the coupling between the 
%two-fold degenerate IR active 
E$_{u}^1$ mode 
and other modes, whose irreducible representation are respectively denoted 
$\Gamma_{E_u}$ and $\Gamma$. Group theory tells us that the modes that are potentially
coupled to the IR modes are such that the symmetrized representation 
$\left[ \Gamma_{E_u} \otimes \Gamma_{E_u} \right]_s$ contains $\Gamma$\cite{radaelli_2018}.
As the point group of Bi$_2$Te$_3$ is D$_{3d}$, it's easy to show that
$\left[ \Gamma_{E_u} \otimes \Gamma_{E_u} \right]_s=A_{1g}\oplus E_g$.
Thus, the energy of the crystal per unit cell, denoted as $V$, is computed as 
a function of $Q_{R,0}$ (coordinate of the $A_{1g}^1$ mode), $Q_{R, i}$ 
(coordinates of the $E_g^1$ mode) and $Q_{IR, i}$ (coordinates of the $E_u^1$ mode), 
where $i=1, 2$ since the $E_g^1$ and $E_u^1$ modes are two-fold degenerate. 
To alleviate the notations, the upper index for both the $E_g^1$ and $E_u^1$ modes 
is suppressed in the following but remains implicit.
The full
computation of $V$ is cumbersome since it depends on five coordinates. However, $V$
must necessarily be invariant by all elements of the point group, leading to a
substantial simplification. Let $Q_{IR, 1}$ and $Q_{IR, 2}$ be the coordinates of
the $E_u$ mode whose polarization respectively lies along one of the two-fold axis 
(x-axis) and one of the mirror plane (y-axis). 
The relevant symmetry elements are shown in Fig. \ref{Structure_fig}(b).
Then, we choose the coordinates 
$Q_{R,1}$ and $Q_{R,2}$ such that they respectively transform as 
$Q_{IR, 2}^2-Q_{IR, 1}^2$ and $Q_{IR, 1} Q_{IR, 2}$. With such a choice, the 
energy $V$ can be splitted into an harmonic part defined as:
\begin{eqnarray}
V_{harm}/M &= &\frac{1}{2} \Omega_{R,0}^2 Q_{R,0}^2 + 
\frac{1}{2} \Omega_{R}^2 Q_{R}^2 +
\frac{1}{2} \Omega_{IR}^2 Q_{IR}^2,
\end{eqnarray}
and an anharmonic part defined as:
\begin{eqnarray}\label{V_anh_def}
V_{anh}/M &= & 
b Q_{IR}^4
+ g_1 \left[Q_{R,1}\left(Q_{IR,2}^2 -Q_{IR,1}^2  \right)  
+ 2 Q_{IR,1} Q_{IR,2} Q_{R,2} \right] \nonumber \\
& & + g_0 Q_{R,0} Q_{IR}^2+ a_0 Q_{R,0}^3 + a_1\left[Q_{R,1}^3 -3 Q_{R,1} Q_{R,2}^2 \right],
\end{eqnarray}
where $Q_{R/IR}^2=\sum_{i=1}^2 Q_{R/IR, i}^2$, 
%$M=\sum_{p=1}^5 M_p$ is the total mass of the unit cell,
$M$ is the total mass of the unit cell,
$\Omega_{R,0}$, $\Omega_{R}$ and
$\Omega_{IR}$ are respectively the angular frequencies of 
the $A_{1g}$, $E_g$ and $E_u$ modes and
$g_0$ ($g_1$) denotes the coupling constant between the $E_u$ modes
and the $A_{1g}$ ($E_g$) modes.

We first consider the energy surface corresponding to $Q_{R,1}=Q_{R,2}=Q_{IR,2}=0$.
In such a case, we have:
\begin{equation}\label{surface1} 
V(Q_{IR, 1}, Q_{R, 0})/M = \frac{1}{2} \Omega_{R,0}^2 Q_{R, 0}^2 
+\frac{1}{2} \Omega_{IR}^2 Q_{IR, 1}^2 
+ b Q_{IR, 1}^4 + g_0 Q_{R, 0}  Q_{IR, 1}^2+ a_0 Q_{R,0}^3.
\end{equation} 
We next consider the energy surface
corresponding to  $Q_{R,0}=Q_{R,2}=Q_{IR,2}=0$. In such a case, we have:
\begin{equation}\label{surface2} 
V(Q_{IR, 1}, Q_{R, 1})/M = \frac{1}{2} \Omega_{R}^2 Q_{R, 1}^2 
+\frac{1}{2} \Omega_{IR}^2 Q_{IR, 1}^2 
+ b Q_{IR, 1}^4 - g_1 Q_{R, 1}  Q_{IR, 1}^2 + a_1 Q_{R,1}^3.
\end{equation} 
For each couple of values $Q_{IR,1}$ and $Q_{R,i}$ ($i=0, 1$), 
the displacements along direction $\alpha$ of each atom $p$ belonging
to the unit cell are given by:
\begin{equation}
{u}_p^\alpha = \sqrt{\frac{M}{M_p}} \sum_{\lambda=IR,1; R,i} Q_\lambda
{\epsilon}_p^\alpha (\lambda),
\end{equation}
where $M_p$ is the mass of atom $p$  and
${\bf{\epsilon}}_p (\lambda)$ is the displacement
of atom $p$ for the zone-center mode $\lambda$.
%With such a definition,
%$Q_\lambda$ has the unit of a length and $Q_{IR,1}$ ($Q_{R,i}$ where $i=0, 1$) 
%is varied between -6 pm (-4 pm) and 6 pm (4pm) with a step of 0.4 pm (0.2 pm).
The two energy surfaces computed at the LDA level for $Q_{IR, 1}$ ($Q_{R,i}$) varying 
between -6 pm (-4 pm) and 6 pm (4 pm) with a step of 0.4 pm (0.2 pm)  
%$V(Q_{IR, 1}, Q_{R, 0})$ and $V(Q_{IR, 1}, Q_{R, 1})$ 
are respectively 
displayed in Fig. \ref{Energy-vs-Q-fig}(a) and Fig. \ref{Energy-vs-Q-fig}(b). 
\begin{figure}[!hbt]
\begin{center}
\vskip0.5truecm
\includegraphics[angle=0, scale=0.3]
{./Fig-6-PRB.eps}
\end{center}
%\vskip -0.75truecm
\caption{
\label{Energy-vs-Q-fig}
$V(Q_{IR,1}, Q_{R,i})$ (in meV per unit cell) where i=0 (left panel) and i=1 (right panel),
as a function of $Q_{R,i}$ (in pm) for different values of $Q_{IR,1}$
ranging from $0$ pm to $6$ pm with a step of $2$ pm.
The circles correspond to the results of LDA calculations and the solid lines
represent the curves arising from a simultaneous
least squares fit of the energy surfaces
$V(Q_{IR,1}, Q_{R,i})$ where $i=0$ ($A_{1g}$ mode) and $i=1$ ($E_g$ mode). The stars
in panel (a) denote the positions of the minima.
}
\end{figure}
As shown in Fig. \ref{Energy-vs-Q-fig}(a), 
the equilibrium value of $Q_{R,0}$ is
slightly displaced towards negative values when $Q_{IR,1}$ increases or decreases
as Eq. \ref{surface1} leads to $Q_{R,0}^{eq}=-g_0 Q_{IR,1}^2/\Omega_{R,0}^2$
when the term $a_0 Q_{R,0}^3$ is neglected.
The unkown coefficients of the full energy surface, that should be  invariant by all
elements of the point group, are obtained from a simultaneous least squares fit of the two
above mentioned energy surfaces and gathered in table \ref{coeff_tab}.
\begin{table}
%\caption{The values of the coefficients of the polynomial for energy
%surfaces of Raman ($A_{1g}$) and IR ($E_{1u}$) modes obtained from a fit
%of the calculated energy surface.} 
\caption{Values of the coefficients of the polynomial extracted from a
simultaneous least squares fit of the two energy surfaces 
$V(Q_{IR, 1}, Q_{R, 0})$ ($E_{u}^1/A_{1g}$) 
and $V(Q_{IR, 1}, Q_{R, 1})$ ($E_{u}^1/E_{g}^1$). Here, $M$ is the mass of the
unit cell and the values in parentheses have been inferred 
from the DFPT calculations\cite{busselez_2023} } 
\label{coeff_tab}
\begin{tabular}{llll}
\hline
$\frac{1}{2} M \Omega_{IR}^2$  (Ha.bohr$^{-2}$) & 0.0433 (0.0442) &
$\frac{1}{2} M \Omega_{R,0}^2$  (Ha.bohr$^{-2}$) &  0.0502 (0.0495)  \\
$\frac{1}{2} M \Omega_{R}^2$ (Ha.bohr$^{-2}$)  &  0.0207 (0.0205) & $M b$ (Ha.bohr$^{-4}$) & 0.2666 \\
$M g_0$ (Ha.bohr$^{-3}$) & 0.1203 & $M g_1$ (Ha.bohr$^{-3}$) & -0.0204 \\
$M a_0$ (Ha.bohr$^{-3}$) &  -0.0085 & $M a_1$ (Ha.bohr$^{-3}$) &  -0.0063 \\
\hline
\end{tabular}
\end{table}
We note that the quadratic terms in the phonon coordinates are in good overall agreement 
with the values inferred from the DFPT results\cite{busselez_2023} and displayed
in parentheses in table \ref{coeff_tab}. It's worth highlighting that the frequencies
of both $E_u^1$ and $A_{1g}^1$ modes, denoted as $\Omega_{IR}$ and $\Omega_{R,0}$,
 are respectively overestimated and underestimated 
from 8\% with respect to the experimental frequencies (See Table \ref{effective_charge_tab}). 
Thus, these
frequencies are slightly renormalized to match the experimental frequencies while the
other 6 parameters shown in table \ref{coeff_tab} are kept unchanged.
%The only exception concerns the $E_u^1$ mode
%which is strongly anharmonic and strongly coupled to the $A_{1g}$ mode as illustrated by the
%non negligible contributions of the $b\left[Q_{IR,1}^4+Q_{IR,2}^4 + 2 Q_{IR,1}^2 Q_{IR,2}^2  \right]$
%and $g_0 Q_{R,0} \left[Q_{IR,1}^2 + Q_{IR,2}^2 \right]$ terms to the potential energy $V$.
%It's also worth poynting out that the $E_u^1$ mode is also slightly coupled to the $E_g$ mode but
%that the contribution of the 
%$g_1 \left[Q_{R,1}\left(Q_{IR,2}^2 -Q_{IR,1}^2  \right) + 2 Q_{IR,1} Q_{IR,2} Q_{R,2} \right]$ 
%term to the potential energy $V$ is ten times smaller than the contribution of the two previous
%terms. Finally, we checked that our symetry analysis of the potential energy surface $V$ is
%correct by considering the energy surface corresponding to $Q_{R,0}=Q_{R,1}=0$ and
%$Q_{IR,1}=Q_{IR,2}=Q$. As shown in Fig. \ref{Energy-vs-QR-fig}(b), the computed energy surface 
%for $Q$ ($Q_{R,2}$) varying between -6 pm (-4 pm) and 6 pm (4 pm) 
%with a step of 0.4 pm (0.2 pm) is perfectly fitted by:

%\begin{equation}\label{surface3} 
%V(Q, Q_{R, 2})/M = \frac{1}{2} \Omega_{E_g^1}^2 Q_{R, 2}^2 
%+ \Omega_{E_{u}^1}^2 Q^2 
%+ 4 b Q^4 + 2 g_1 Q_{R, 2}  Q^2
%\end{equation} 
%with the values taken from table \ref{coeff_tab}, assessing the validity of our symetry analyis.

\section{Phonon dynamics including phonon-phonon coupling}\label{Eq_phonon_dynamics_part}
The equation of motion for atom $p$ of mass $M_p$ reads:
\begin{equation}\label{eq_motion}
M_p \ddot{u}_p^\alpha =
- \sum_{p^\prime, \alpha^\prime} 
C_{p, p^\prime}^{\alpha, \alpha^\prime} u_{p^\prime}^{\alpha^\prime}
+ e \sum_\beta Z^*_{p, \beta, \alpha} E^\beta,
\end{equation}
where $u_p^\alpha$ is 
the displacement of atom $p$ ($p \in \{1, \cdots, n\}$ where $n=5$) along the direction $\alpha$
and $E^\beta$ the component of the macroscopic field along direction $\beta$ with
$\alpha, \beta \in \{1,2,3\}$. Here, we introduced the elastic constants 
$C_{p, p^\prime}^{\alpha, \alpha^\prime}$ and the dimensionless  
Born effective charge tensors $Z^*_{p, \beta, \alpha}$,
respectively defined by:
\begin{equation}
C_{p, p^\prime}^{\alpha, \alpha^\prime} = 
\frac{\partial^2 E_{tot}}{\partial u_p^\alpha \partial u_{p^\prime}^{\alpha^\prime}},~
\textrm{and}~
Z^*_{p, \beta, \alpha}=\frac{1}{e} \frac{\partial F_p^\alpha}{\partial E^\beta},
\end{equation}
where $E_{tot}$ is the total energy per unit cell 
and $F_p^\alpha$ is the component of the force along direction $\alpha$ 
acting on atom $p$. The dynamical matrix at the zone center ($3n \times 3n$ matrix),
defined as $D_{p, p^\prime}^{\alpha, \alpha^\prime}=C_{p, p^\prime}^{\alpha, \alpha^\prime}/\sqrt{M_p M_{p^\prime}}$, is easily diagonalized:
\begin{equation}\label{eigen_problem}
\sum_{p^\prime, \alpha^\prime} D_{p, p^\prime}^{\alpha, \alpha^\prime}
\epsilon_{p^\prime}^{\alpha^\prime}(\lambda) = \omega_\lambda^2 \epsilon_{p}^{\alpha}(\lambda).
\end{equation}
Here, $\omega_\lambda$ 
and $\epsilon_{p}^{\alpha}(\lambda)$ are respectively the frequency and the displacement of
atom $p$ along $\alpha$ for the mode $\lambda$, where $\lambda \in \{1, \cdots, 3n\}$.
The eigenvectors of Eq. \ref{eigen_problem} satisfy the orthogonality relations:
\begin{equation}\label{orthogonality}
\sum_{p, \alpha} \epsilon_{p}^{\alpha}(\lambda) \epsilon_{p}^{\alpha}(\lambda^\prime) =
\delta_{\lambda, \lambda^\prime}, 
\end{equation}
since the zone center dynamical matrix is real and symmetric. Plugging
\begin{equation}
u_p^\alpha = \sqrt{\frac{M}{M_p}} \sum_{\lambda^\prime} Q_{\lambda^\prime} 
\epsilon_p^\alpha(\lambda^\prime)
\end{equation}
into Eq. \ref{eq_motion}, multiplying both members of Eq. \ref{eq_motion} by 
$\epsilon_p^\alpha(\lambda)$ and summing over $p$ and $\alpha$ leads to:
\begin{equation}\label{simplified_ed}
\ddot{Q}_\lambda + \omega_\lambda^2 Q_\lambda = \frac{e}{M} \sum_{\beta} Z_\beta(\lambda) E^\beta, 
\end{equation}
where Eq. \ref{eigen_problem} and Eq. \ref{orthogonality} are used. Here, 
the mode effective charge $Z_\beta(\lambda)$ defined as:
\begin{equation}\label{def_X}
Z_{\beta} (\lambda) = \sum_{p, \alpha} \sqrt{\frac{M}{M_p}} \epsilon_p^\alpha(\lambda)
Z^*_{p, \beta, \alpha},
\end{equation} 
is non zero for ungerade modes. 
Both the $E_u^1$ and $E_u^2$ modes, which couple to an electric field 
polarized perpendicular to the trigonal axis, exhibit a two-fold degeneracy. 
For each of these two modes, 
the $IR,1$ mode is chosen to be polarized along the x-axis (along one of the three $C_2$ axis
shown in Fig. \ref{Structure_fig}(b)) 
while the $IR,2$ mode is polarized along the y-axis (in one of the three $\sigma_d$ planes
shown in Fig. \ref{Structure_fig}(b)).
Thus, the effective charges for these  modes read:
\begin{equation}\label{def_Z_Eu}
Z_{\beta} (IR, i) = \delta_{\beta, i} Z_\perp ~\textrm{where}~
Z_\perp = \sum_{p} \sqrt{\frac{M}{M_p}} \epsilon_p^i(IR, i)
Z^*_{\perp}(p)
\end{equation} 
exhibit no dependence on $i$.
For the sake of completeness, the frequencies, mode effective
charges and normalized atomic displacements are displayed in Table \ref{effective_charge_tab}.
\begin{table}[!hbt]
\caption{Frequencies (in THz), mode effective charges (dimensionless quantities)
and normalized atomic displacements (see Eq. \ref{orthogonality}) 
for the zone center modes of Bi$_2$Te$_3$. The calculations have been done for the experimental
lattice structure\cite{francombe_1958} using the values of the Born effective charges
reported in our previous work\cite{busselez_2023}. Both experimental Raman\cite{Wang_2013}
and Infrared\cite{richter_1977} frequencies measured at 300 K are indicated in parentheses.
}
\label{effective_charge_tab}
\begin{tabular}{lccccccccc}
\hline
Symmetry & Frequency &\multicolumn{2}{c}{Mode effective charge} &\multicolumn{5}{c}{Normalized atomic displacements} & Direction\\
         &           & $Z_\perp$  & $Z_\parallel$ &Te$_1$ & Bi & Te$_2$ & Bi & Te$_1$ &  \\
\hline
E$_g^1$  & 1.10 (1.11) &  -   & - & 0.369 & -0.603 & 0.0 & 0.603 & -0.369 & $x$, $y$\\
A$_{1g}^1$  & 1.71 (1.86) &  -   & - & 0.497 & -0.503 & 0.0 & 0.503 & -0.497 & $z$ \\
E$_g^2$  & 2.93 (3.06) &  -   & - &  0.603 &  0.369 & 0.0 &  -0.369 &  -0.603 & $x$, $y$\\
A$_{1g}^2$  & 3.87 (4.05) &  -   & - & 0.503 & 0.497 & 0.0 & -0.497 & -0.503 & $z$ \\
E$_u^1$  & 1.62 (1.50 $\pm$0.06)&  -36.31 &  -  & 0.310&-0.475 & 0.596& -0.475 & 0.310 & $x$, $y$\\
E$_u^2$   & 2.75 (2.82 $\pm$ 0.12)&  -3.93 &   -  & 0.494& -0.114& -0.696& -0.114& 0.494& $x$, $y$\\
A$_{2u}^1$ & 2.79 (2.85 $\pm$ 0.15) &  - &   23.25 & 0.121& 0.262& -0.913& 0.262& 0.121& $z$\\
A$_{2u}^2$ & 3.60  (3.60 $\pm$ 0.15) &  - &  -16.54 & 0.571& -0.413& -0.085& -0.413& 0.571& $z$\\
\hline
\end{tabular}
\end{table}
We observe that the $E_u^1$ mode is strongly coupled to an in-plane electric field
as it exhibits the largest effective charge ($Z_\perp \sim -36.3$).
It's worth pointing out that the sign
of $Z_\perp$ depends on the phase choice ($\pm 1$) for the real eigenvectors of the
zone center dynamical matrix. In our case, $Z_\perp$ is negative in accordance with
the fact that $Q_{IR, 1}$ ($Q_{IR, 2}$) should decrease when the electric field is polarized
along $+x$ ($+y$) since the Te (Bi) atoms behave respectively as negatively (positively)
charged ions.

Eq. \ref{simplified_ed} assumes that the atoms undergo undamped harmonic oscillations around
their equilibrium positions. Going beyond the harmonic approximation and taking into 
account the finite lifetime of the phonons leads to:
\begin{equation}\label{eq_motion2}
\ddot{Q}_\lambda + 2 \gamma_\lambda \dot{Q}_\lambda + \omega_\lambda^2 Q_\lambda 
+\frac{1}{M} \frac{\partial V_{anh} }{\partial Q_\lambda  }= 
\frac{e}{M} \sum_{\beta} Z_\beta(\lambda) E^\beta, 
\end{equation}
where $\gamma_\lambda=1/\tau_\lambda$ is the phonon inverse lifetime and where
$V_{anh}$ is given by Eq. \ref{V_anh_def}.
We can write the equations of motion for the three modes 
%($E_u^1 \equiv IR,i / A_{1g}^1 \equiv R,0 /E_g^1 \equiv R,i$, where $i=1, 2$) 
involved in the dynamics following the arrival of the experimental THz pulse polarized 
perpendicular to the trigonal axis. Indeed,
using Eq. \ref{eq_motion2} and Eq. \ref{V_anh_def}, we obtain: 
%\begin{eqnarray}
\begin{equation}\label{eq_motion1}
\left\{
\begin{array}{l}
\ddot{Q}_{IR, 1} + 2 \gamma_{IR} \dot{Q}_{IR, 1}
+\left[\Omega_{IR}^2 + {\mathbf{4 b Q_{IR}^2}}+ {\mathbf{2(g_0 Q_{R,0} -g_1 Q_{R,1})}} \right] Q_{IR, 1}  
+ {\mathbf{2 g_1 Q_{IR, 2} Q_{R, 2}}}=\frac{F^1(t)}{M}  \\
\ddot{Q}_{IR, 2} + 2 \gamma_{IR} \dot{Q}_{IR, 2}
	+ \left[\Omega_{IR}^2 + {\mathbf{4 b Q_{IR}^2}}+ {\mathbf{2(g_0 Q_{R,0} +g_1 Q_{R,1})}} \right] Q_{IR, 2} 
	+ {\mathbf{2 g_1 Q_{IR, 1} Q_{R, 2}}}=\frac{F^2(t)}{M}  \\
\ddot{Q}_{R, 0} + 2 \gamma_{R,0} \dot{Q}_{R, 0}
+ \Omega_{R,0}^2 Q_{R, 0} 
= - g_0 Q_{IR}^2 -{\mathbf{3 a_0 Q_{R,0}^2}}  \\
\ddot{Q}_{R, 1} + 2 \gamma_{R} \dot{Q}_{R, 1}
+ \Omega_{R}^2 Q_{R, 1} 
= - g_1 \left[Q_{IR,2}^2-Q_{IR,1}^2\right] 
-{\mathbf{3 a_1 \left[Q_{R, 1}^2-Q_{R, 2}^2 \right]}} \\
\ddot{Q}_{R, 2} + 2 \gamma_{R} \dot{Q}_{R, 2}
+ \Omega_{R}^2 Q_{R, 2} 
= - 2 g_1 Q_{IR,1} Q_{IR,2} + {\mathbf{6 a_1 Q_{R, 1} Q_{R, 2}}},
%\frac{e}{M} Z_\perp E_1 \equiv \frac{F_1(t)}{M}
%\end{eqnarray}
\end{array}\right.
\end{equation}
where $Q_{IR}^2=Q_{IR,1}^2+Q_{IR,2}^2$ and $F^\alpha(t)=e Z_\perp E^\alpha(t)$
with $\alpha=1, 2$.
%with $E^\alpha(t)$ representing the component along direction $\alpha$ 
%of the THz electric field inside the Bi$_2$Te$_3$ nanofilm. 
The components of the THz pulse inside the Bi$_2$Te$_3$ nanoscale thin film are given by:
\begin{equation}\label{Electric_field_component}
E^\alpha(t)=p E_0(t) \left[ \cos(\Phi) \delta_{\alpha,1} + \sin(\Phi) \delta_{\alpha,2} \right],
\end{equation}
where $E_0(t)$ is described in section \ref{expt-part} (See Eq. \ref{E0_vs_t}),
$\Phi$ is the angle between the electric field and the x-axis 
and $p\sim 2/(1+n_{\textrm{mica}}) \sim 0.57$
since $n_{\textrm{mica}}\sim 2.5$ for $\omega_0/2\pi \sim 0.6$ THz\cite{Janek_2009}.
%It's worth outlining that the electric field inside the nanoscale thin film is
%, as a first approximation, 
%neither dependent on the nanofilm thickness nor on the dielectric properties 
%of the nanofilm as the THz pulse wavelength is much larger than the film thickness.
%Notably, the electric field within the nanoscale thin film is independent of
It's worth outlining that the electric field within the nanoscale thin film is independent of
both the film thickness and its dielectric properties, as
the THz pulse wavelength is far greater than the film thickness.
%the THz pulse wavelength significantly exceeds the film thickness.

\section{Dynamics of the modes for the experimental THz pulse}\label{Dynamics_modes_part}
All parameters entering the equations of motion
%, with the exception of parameters $\gamma_{IR}$, $\gamma_{R,0}$ 
%and $\gamma_{R}$, 
are gathered in
table \ref{coeff_tab} and \ref{effective_charge_tab}.
The only unknown parameters are $\gamma_{IR/R,i}$ that are related to the inverse phonon
lifetimes. We choose $\gamma_{IR}\sim 0.44$ ps$^{-1}$ as this value allows to reproduce
the IR spectra of Bi$_2$Te$_3$\cite{busselez_2023}
and consider that $\gamma_{R,0}\sim 0.28$
ps$^{-1}$ as the full width at half maximum of the A$_{1g}$ mode is $\sim 3$ cm$^{-1}$ at
room temperature\cite{vilaplana_2011}. As no data is available for the $E_g$ mode, we
assume that the inverse phonon lifetime for this mode is the same than for the $A_{1g}$ mode.
We solve the set of coupled differential equations (see Eq. \ref{eq_motion1}) using 
a fourth order Runge-Kutta scheme with a time step of one fs. 
%When using the characteristics
%of the experimental THz pulse ($E_0=340$ kV.cm$^{-1}$, $\sigma_0=0.79$ ps, $\Phi_0=0.117\pi$
%and $\omega_0/2\pi=0.64$ THz), our simulations show that the feedback effect of the Raman
%active modes on the IR active mode as well as the renormalization of the IR mode 
%($4b Q_{IR}^2 \ll \Omega_{IR}^2$) can be safely neglected (See the terms colored in red in
%Eq. \ref{eq_motion1}). Furthermore, the terms colored in green in Eq. \ref{eq_motion1}
%can also be ignored. Therefore, the equations of motion can be simplified 
%(See Eq. 2 of the main text) provided that the experimental value of $E_0$ is
%considered.
Interestingly, our simulations show that the dynamics of the modes are not affected 
when the bold terms in Eq. \ref{eq_motion1} are neglected. Thus, solving the following
simplified equations:
\begin{equation}\label{eq_motion_simplified}
\left\{
\begin{array}{l}
\ddot{Q}_{IR, \alpha} + 2 \gamma_{IR} \dot{Q}_{IR, \alpha}
+\Omega_{IR}^2 Q_{IR, \alpha}  =F^\alpha(t)/M \\
\ddot{Q}_{R, i} + 2 \gamma_{R,i} \dot{Q}_{R, i}
+ \Omega_{R,i}^2 Q_{R, i}
= F_{R,i}^{anh}/M,
\end{array}\right.
\end{equation}
for $\alpha=1,2$ and $i=0,1,2$ and 
where $\Omega_{R,1}=\Omega_{R,2}=\Omega_{R}$, 
$\gamma_{R,1}=\gamma_{R,2}=\gamma_{R}$,
yields accurate results for the dynamics of
the three modes provided that the experimental field amplitude discussed 
in section \ref{expt-part} is considered. Here, the forces are respectively given by:
\begin{equation}\label{F_R0_anh}
F_{R,0}^{anh}=-M g_0 \left[Q_{IR,1}^2+Q_{IR,2}^2\right]
\end{equation}
and:
\begin{equation}\label{F_R1_anh}
F_{R,i}^{anh}=-M g_1 \left[\delta_{i,1}\left(Q_{IR,2}^2-Q_{IR,1}^2\right) +
2 \delta_{i,2} Q_{IR,1} Q_{IR,2}\right].
\end{equation}
Note that the forces resulting from two-photon direct excitation
of the Raman active modes\cite{maehrlein_2017, juraschek_2018, courtney_2019}, 
as well as those arising from the redistribution 
of free electrons induced by the THz pulse, are neglected 
(See section \ref{Raman_forces_part}).
\begin{figure}[!hbt]
\begin{center}
\vskip0.9truecm
\includegraphics[angle=0, scale=0.3]
{./FIG-7-PRB.eps}
\end{center}
%\vskip -0.75truecm
\caption{
\label{dynamics-fig}
Calculated time evolution of the E$_{u}^1$ (a), 
A$_{1g}^1$ (b) and E$_{g}^1$ (c) mode coordinates following
the arrival of a THz pulse polarized along the x-axis ($\Phi=0$) with an amplitude
$\sim$ 193 kV.cm$^{-1}$ inside the Bi$_2$Te$_3$ nanofilm ($E_0 \sim 340$ kV.cm$^{-1}$).
The solid lines represent $Q_{IR,1}$, $Q_{R,0}$ and $Q_{R,1}$ for 
$\Phi_0=0.117 \pi$ while the dashed lines represent $Q_{IR,1}$, $-Q_{R,0}$ and
$-Q_{R,1}$ for $\Phi_0=0.117 \pi + \pi/2$.
The grey shaded areas represent the envelope of the THz pulse. 
The normalized Fourier Transforms (FT) of the phonon coordinates $Q$, denoted as $\hat{Q}$, 
are shown in panels (d), (e) and (f) for $\Phi_0=0.117 \pi$.
The FT of the THz electric field is shown as a dashed line in panel (d).
%The
%thin black (dashed red) curves are obtained by solving Eq. \ref{equation_motion} 
%when only the anharmonic forces (Raman forces) are
%taken into account while the thick blue curves include both forces.
%The THz pulse enveloppe is schematically shown as the cyan filled curve.
%Note that the anharmonic forces arise from the direct driving of the $E_u^1$ mode
%shown in pannel (a).
%The inset shows the temporal evolution
%of the $\sim$ 210 kV.cm$^{-1}$ experimental THz pulse inferred from an
%electro-optic sampling technique (black circles) as well as the fitting curve (red curve).
%(b) Calculated time evolution of the A$_{1g}$ mode coordinate $Q_{R,0}$ (in pm).
%(c) Calculated time evolution of the E$_{g}^1$ mode coordinate $Q_{R,1}$ (in pm).
%(a) Calculated time evolution of the E$_{1u}$ mode coordinate $Q_{IR,1}$ (in pm) following
%the arrival of a Thz pulse polarized along the x-axis with an amplitude
%$\sim$ 170 kV.cm$^{-1}$ inside the Bi$_2$Te$_3$ nanofilm.
%The inset shows the temporal evolution
%of the experimental Thz pulse (in kV.cm$^{-1}$) inferred from an
%electro-optic sampling technique (black curve) as well as the fitting curve (red curve).
%(b) Calculated time evolution of the A$_{1g}$ mode coordinate $Q_{R,0}$ (in pm).
%(c) Calculated time evolution of the E$_{g}^1$ mode coordinate $Q_{R,1}$ (in pm).
}
\end{figure}
In Fig. \ref{dynamics-fig}, we report  the dynamics of the $E_u$, $A_{1g}$ and $E_g$ modes 
for a THz pulse polarized along the x-axis.
% with an amplitude $\sim 193$ kV.cm$^{-1}$ inside the Bi$_2$Te$_3$ nanofilm. 
As shown in panel (a), the $Q_{IR,1}$ coordinate 
is phase shifted from $\pi$ $(Z_\perp <0)$ with respect to the
THz pulse shown in Fig. \ref{fig2}(a) and
displays tiny oscillations at the frequency of the $E_u$ mode that are
seen as a small bump in the Fourier Transform (FT) of $Q_{IR,1}$, denoted as $\hat{Q}_{IR,1}(\nu)$
and shown in panel (d).
This  behaviour can be readily understood since the THz pulse is off-resonance with
the IR mode (See Fig. \ref{fig2}(b)).
%the long time behaviour of $Q_{IR,1}$ 
%is analytically given\cite{SM} by
%$Q_{IR,1}^{long}(t) \sim \overline{Q} \exp[-\gamma_{IR} t] 
%\exp[-\sigma^2 \Delta\omega^2/4] \cos[\tilde{\omega}t+\Psi]$,
%where $\Delta \omega=\tilde{\omega}-\omega_0$, 
%$\overline{Q}=-e Z_\perp E_0 \sigma \sqrt{\pi} \exp[\sigma^2 \gamma_{IR}^2/4]/(2M\tilde{\omega})$,
%$\Psi=\Phi_0-\gamma_{IR} \sigma^2 \Delta \omega/2$ and
%$\tilde{\omega}=\sqrt{\Omega_{IR}^2-\gamma_{IR}^2}$. As the THz pulse is non-resonant with
%the IR mode ($\Delta \omega \gg 2/\sigma$), the oscillation amplitude is very small
%for $t > 2 \sigma$. 
As shown in Fig. \ref{dynamics-fig}(b), the symmetry preserving 
$Q_{R,0}$ mode oscillates
around a displaced equilibrium position during the THz pulse and oscillates around its 
equilibrium position afterward. The Fourier transform of the $Q_{R,0}$ mode, denoted as
$\hat{Q}_{R,0}(\nu)$ and shown in panel (e), displays a peak at 1.86 THz, corresponding to 
the natural frequency oscillation of the $A_{1g}$ mode. Reminding that the driving force
acting on the $A_{1g}$ mode is given by Eq. \ref{F_R0_anh} where $Q_{IR,2}=0$,
the $A_{1g}$ mode is not efficiently driven.
%E as the double of the THz frequency pulse does not
%match the frequency of the $A_{1g}$ mode. 
%The maximum value of $\left|Q_{R,0}\right|$ is $0.06$ pm,
%while reaching pm when the central frequency of the THz pulse is doubled (see SM). 
Indeed, the maximum value of $\left|Q_{R,0}\right|$ is $0.06$ pm.
%for $\omega_0/2\pi=0.64$ THz,
%while reaching 0.14 pm when $\omega_0/2\pi$ is tuned to half the frequency of the
%$A_{1g}$ mode\cite{SM}.
%$\Omega_{A_{1g}^1}$ (see SM). 
It's also important to remark that the dynamics of the $A_{1g}$ mode, unlike the dynamics 
of the $E_g$ mode, does not depend on the polarization of the THz field.
As shown 
in Fig. \ref{dynamics-fig}(c), the behaviour of the $Q_{R,1}$ coordinate is similar to
the behaviour of the $Q_{R,0}$ coordinate. The main difference is that the maximum value
of  $\left|Q_{R,1}\right|$ is $0.018$ pm.
% while reaching pm when the frequency of the THz pulse is doubled. 
Here, the driving force given by Eq. \ref{F_R1_anh} for $i=1$ and $Q_{IR,2}=0$,
%$F_{R,1}^{anh}=M g_1 Q_{IR,1}^2$ ($Mg_1 \sim -0.02$ Ha$\cdot$bohr$^{-3}$) 
is six times smaller than $F_{R,0}^{anh}$ but the double 
of the frequency of the THz pulse fortuitously matches the frequency of the $E_g$ mode.
It's worth outlining that only the $Q_{R,1}$ mode is coherently driven when the THz pulse
is polarized along the $x$-axis ($\Phi=0$) or $y$-axis ($\Phi=\pi/2$) 
while only the $Q_{R,2}$ mode is driven
when $\Phi=\pi/4, 3\pi/4$. 
Thus, the polarization of the THz pulse offers the unique opportunity
to steer the dynamics of the $E_g$ mode and especially to 
transiently lower the crystal symmetry to $2/m$ ($\overline{1}$) when $Q_{R,1}$ 
($Q_{R,2}$) is coherently driven, as discussed in section \ref{hypothetical_pulse_part}
for an hypothetical THz pulse resonant with the $E_u^1$ mode. 
%Interestingly, the symmetry lowering is 
%enhanced and lasts longer than $\sigma\sim 0.8$ ps when the THz pulse is resonant with 
%the $E_u$ mode\cite{SM}.

It's also tempting to play 
with the carrier envelope phase $\Phi_0$ of the THz pulse (See Eq. \ref{E0_vs_t}). 
The maximum values of the 
phonon coordinates are only slightly affected by $\Phi_0$ ruling out the idea to adjust
$\Phi_0$ in order to enhance the amplitudes of the $A_{1g}$ and $E_g$ modes. However, it's
easy to play with the phases of $Q_{R,i}$, where $i=0,1,2$. We expect the long time
dynamics of these modes to be well described by 
$Q_{R,i}^{long}(t, \Phi_0)=A_{R,i} \cos\left[\tilde{\Omega}_{R,i}t + f_i(\Phi_0)\right]$,
where $\tilde{\Omega}_{R,i}=\sqrt{\Omega_{R,i}^2 - \gamma_{R,i}^2}$ is the
renormalized angular phonon frequency and $f_i(\Phi_0\pm\pi)=f_i(\Phi_0)\pm 2\pi$. Indeed, a change in the sign of the THz
pulse does not affect the long time dynamics of the $Q_{R,i}$ modes as the driving 
forces remain unchanged. The only way to satisfy the previous equation is to impose 
$f_i(\Phi_0)=2\Phi_0+\Phi_{R,i}$. This property has been checked in our numerical calculations.
As shown in Fig. \ref{dynamics-fig}, the long time dynamics of the $A_{1g}$ and $E_g$
phonons corresponding to $\Phi_0=0.117 \pi + \pi/2$ (dashed curves) are phase shifted
from $\pi$ with respect to the dynamics corresponding to $\Phi_0=0.117 \pi$ (solid curves).
Interestingly, the coherent phonon dynamics in diamond arising from a THz sum-frequency 
process displays a similar dependence on the carrier envelope phase\cite{maehrlein_2017}. 

\section{Simulation of the detection process}\label{detection_part}
\begin{figure}[!hbt]
\begin{center}
\vskip0.5truecm
\includegraphics[angle=0, scale=0.3]
{./FIG-8-PRB.eps}
\end{center}
%\vskip -0.75truecm
\caption{
\label{optics-fig}
(a) Computed real (dashed lines) and imaginary part (thin lines) 
of the dielectric function $\epsilon_\perp$
at the RPA level for an electric field perpendicular to the
trigonal axis and for $Q_{R,0}=\pm 1.2$ pm as 
a function of the photon energy (in eV).
%(b) Real (blue dashed line) and imaginary part (black line) 
%of the Raman tensor $\partial \chi_\perp / \partial Q_{R,0}$ 
%(in pm$^{-1}$) as a function of the photon energy (in eV) computed with a 5-point 
%stencil method where the spacing between points is $0.4$ pm.
(b) Transmittance $\Delta T/T=[T(Q_{R,0})-T(0)]/T(0)$ 
at $400$ nm (wavelength of the probe pulse) 
computed using a transfer matrix approach\cite{katsidis_2002} for the 
heterostructure schematically depicted in the inset: The oxide layer, 
Bi$_2$Te$_3$ film and mica substrate are respectively $4$ nm, $12.2$ nm 
and $17.1$ $\mu$m thick (See section \ref{Sample_part}). 
The phonon coordinate $Q_{R,0}$ is varied between 
-1.2 pm and 1.2 pm with a step of 0.4 pm.
(c) $\Delta T/T$ (dashed blue curve) and $\overline{\Delta T/T}$ (thick blue curve) 
as a function of the time delay $t$ between the probe pulse and 
the pump THz pulse compared to the oscillatory part (open circles) of the transient 
measured transmittance 
%extracted from the raw signal\cite{SM} 
shown in Fig.\ref{fit_fig}(b) as a blue dotted line. 
The experimental value of the THz electric field $E_0 \sim 340$ kV.cm$^{-1}$
is used in the simulations (see section \ref{expt-part}) 
and the measured signal is shifted to match the
theoretical prediction as the zero time delay is ill-defined in our experiments.
% and $\Delta T/T$ or $\overline{\Delta T/T}$ do not depend 
%on the angle $\Phi$ between the electric field and the $x$-axis.
}
\end{figure}

It is important to simulate the detection process in order to assess the 
validity of our approach and confirm that our model captures the most relevant
mechanisms. Indeed, we could expect that the optical properties of the 
Bi$_2$Te$_3$ nanoscale thin film are modulated by the coherently driven $A_{1g}$ mode,
leading to a variation of the measured transmittance of the heterostructure.
Both the computed real and imaginary part of the dielectric function 
$\epsilon_\perp(\omega)$ at the Random Phase Approximation (RPA) 
level\cite{yambo_2009, yambo_2019, arnaud_2001, 
busselez_2023} for $Q_{R,0}=\pm 1.2$ pm
are displayed in Fig. \ref{optics-fig}(a). As $Q_{R,0}$ increases, 
$\textrm{Im}\left[\epsilon_\perp(\omega)\right]$ decreases when $\hbar \omega \leq 1.5$ eV.
Thus, the oscillator strength is transferred to the high energy side as the optical sum rule:
\begin{equation}
\int_0^\infty d(\hbar \omega) ~\hbar \omega \textrm{Im}\left[\epsilon_\perp(\omega)\right] =
\frac{\pi}{2} \left(\hbar \omega_p\right)^2,
\end{equation}
should be obeyed. 
Here, the plasmon energy $\hbar \omega_p \sim 15.1$ eV depends
only on the number of valence electrons per unit cell but not on $Q_{R,0}$. 
The fact that $\textrm{Im}\left[\epsilon_\perp(\omega)\right]$ strongly depends on $Q_{R,0}$
does not arise from the change of the joint density of states but from the change
in matrix elements of the dipolar operator when the atoms are displaced along the
	$A_{1g}$ phonon mode. 
When $Q_{R,0}$ varies from $-1.2$ pm to $1.2$ pm,
the imaginary part (real part) of $\epsilon_\perp$ varies from $10.60$ ($-9.83$) to
$11.24$ ($-9.71$) for $\hbar \omega=3.1$ eV. Hence, the change in the optical index
of Bi$_2$Te$_3$ at the probe wavelength is essentially related to the change in
the imaginary part of the dielectric constant.

%Our measurements are carried out on a Bi$_2$Te$_3$ nanofilm deposited on
%a mica substrate\cite{weis_2015, Levchuk_2020, SM}. The nanofilm is covered with an oxidized
%layer whose exact nature remains unknown.
%However, X-ray photoelectron spectroscopy experiments\cite{weis_2015}
%have shown the presence of Bi-O and Te-O bonds that are the hallmarks of
%an oxidized layer displaying the properties of a glass\cite{bando_2000}
%which might share some properties with bulk Bi$_2$O$_3$ and bulk TeO$_2$.
As discussed in section \ref{Sample_part}, the heterostructure studied 
experimentally is rather complex but can be characterized in terms of
layer thicknesses, with some uncertainty regarding the true nature of the
oxide layer.
Thus, we consider an hypothetical structure schematically depicted in the
inset of Fig. \ref{optics-fig}(b) with an oxidized layer either made 
of Bi$_2$O$_3$ or TeO$_2$. Using a transfer matrix approach\cite{katsidis_2002} 
and the complex refractive index of Bi$_2$Te$_3$ evaluated 
at the RPA level\cite{busselez_2023} (See Fig. \ref{Optical_index_fig}(a))
as well as the experimental optical index of Bi$_2$O$_3$\cite{dolocan_1981},
TeO$_2$\cite{uchida_1971} and mica\cite{nitsche_2004} (See Fig. \ref{Optical_index_fig}(b)), 
the relative transmittance $\Delta T/T$ 
of both envisionned heterostructures is computed as a function of $Q_{R,0}$ when
the mica layer is treated as an incoherent layer. While the transmittance $T$ is
slightly higher for a Bi$_2$O$_3$ layer than for a TeO$_2$ layer, the relative
transmittance is almost unchanged demonstrating that the very nature of the 
oxide layer is unimportant. As shown in Fig. \ref{optics-fig}(b), 
$\Delta T/T=\beta Q_{R,0}$, where $\beta=-1.85$ pm$^{-1}$. Because of 
the finite duration of the probe pulse, 
%and the presence of the mica substrate,
what is measured is not $\Delta T/T=\beta Q_{R,0}(t)$ (see the dashed line
in Fig. \ref{optics-fig}(c)) but rather 
$\overline{\Delta T/T}= \beta Q_{R,0} \otimes g_\sigma(t)$, where the
phonon coordinate $Q_{R,0}$ is convoluted with a normalized Gaussian function 
defined by 
%$g_\sigma(\tau)=\exp\left[-(\tau + \tau_0)^2/\sigma^2\right]/\sqrt{\pi \sigma^2}$. 
$g_\sigma(t)=\exp\left[-t^2/\sigma^2\right]/\sqrt{\pi \sigma^2}$. 
Here, the full width at half maximum (FWHM) is 
$2\sigma \sqrt{\ln 2} \sim 165$ fs.
% and $\tau_0$ is the propagation time 
%of the probe pulse across the mica substrate defined as 
%$\tau_0=d_{mica}/v_g \sim 0.097$ ps, where
%$v_g  \sim 0.59$c is the group velocity\cite{nitsche_2004}. 
As shown in Fig. \ref{optics-fig}(c), the agreement between theory (thick blue line)
and experiment (open circles) is noteworthy especially for time delays larger than
$\sigma \sim 0.8$ ps. The weak discrepancy between theory and experiment for shorter time
delays might be due to the imperfect fit of the oscillatory part of the signal 
(See Fig. \ref{fit_fig}(a)) or to mechanisms not included in our model.

\section{Phonon dynamics including Raman forces}\label{Raman_forces_part}
Up to now, we focused on the nonlinear coupling between the $E_u^1$ modes and
the Raman active modes ($A_{1g}^1$ and $E_g^1$) and ruled out the possibility 
for the Raman active modes to be driven by Raman sum frequency processes which
have been identified as being relevant in diamond\cite{maehrlein_2017} and 
BiFeO$_3$\cite{juraschek_2018}. Let's discuss how the Raman forces can be computed.
The potential energy per unit cell at time $t$ reads:
\begin{equation}\label{coupling}
U(t)=-\frac{v}{2}  \sum_{\alpha} P^\alpha(t) E^\alpha(t),
\end{equation}
where $v$ is the unit cell volume, $E^\alpha$ is the component along $\alpha$ of the THz field inside 
the Bi$_2$Te$_3$ nanofilm and where the component of the polarization along
$\alpha$ reads 
\begin{equation}
P^\alpha(t) = \epsilon_0 \sum_{\beta} 
\int_{-\infty}^t \chi_{\alpha, \beta}(t-t^\prime) E^\beta(t^\prime) dt^\prime.
\end{equation}
Here, $\chi_{\alpha, \beta}(t)$ is the electronic dielectric susceptibility (causal quantity).
Thus, the Raman force acting
on mode $R,i$ ($i=0, 1, 2$) reads:
\begin{equation}\label{Raman_force}
F_{R,i} = - \frac{\partial U(t)}{\partial Q_{R, i}} = 
	\frac{\epsilon_0 v}{2}
\sum_{\alpha, \beta} 
E^\alpha(t) \int_{-\infty}^t 
\frac{\partial \chi_{\alpha, \beta}(t-t^\prime)}{\partial Q_{R,i}} E^\beta(t^\prime) dt^\prime.
\end{equation}
The time dependent Raman tensor $\frac{\partial \chi_{\alpha, \beta}(t)}{\partial Q_{R,i}}$
can be Fourier transformed and the frequency dependent Raman tensors 
$\frac{\partial \hat{\chi}_{\alpha, \beta}(\omega)}{\partial Q_{R,i}}$ have the following
matrix form in our crystallographic axis:
\begin{equation}\label{tensor_R}
Q_{R,0} ~(\textrm{A}_{1g})=\left( \begin{array}{ccc} \hat{a} & 0 & 0 \\
0 & \hat{a} & 0 \\
0 & 0 & \hat{b} \end{array} \right)~,~
Q_{R,1} ~(\textrm{E}_{g})=\left( \begin{array}{ccc} \hat{c} & 0 & 0 \\
0 & -\hat{c} & \hat{d} \\
0 & \hat{d} & 0 \end{array} \right)~,~
Q_{R,2} ~(\textrm{E}_{g})=\left( \begin{array}{ccc} 0 & -\hat{c} & -\hat{d} \\
-\hat{c} & 0 & 0 \\
-\hat{d} & 0 & 0 \end{array} \right).
\end{equation}
Using Eq. \ref{Raman_force} and Eq. \ref{Electric_field_component},
%and giving the fact that:
%\begin{equation}
%E^\alpha(t)=p E_0(t) \left[ \cos(\Phi) \delta_{\alpha,1} + 
%\sin(\Phi) \delta_{\alpha,2} \right],
%\end{equation}
%where $E_0(t)$ is the time profile of the experimental THz pulse 
%(see Eq. \ref{E0_vs_t}), $p$ is a dimensionless parameter accounting
%for the renormalization of the electric field inside the Bi$_2$Te$_3$ nanofilm and 
%$\Phi$ is the angle between the electric field and the $x$-axis, 
it's straightforward to show that the force acting on the
$R,0$ mode ($A_{1g}$ mode) reads:
\begin{equation}\label{F_R0_Raman}
	F_{R,0}(t)=\frac{v \epsilon_0}{2}  p^2 E_0(t) 
\int_{-\infty}^t a(t-t^\prime) E_0(t^\prime) dt^\prime  
\end{equation}
while the forces acting on the $R,1$ and $R,2$ modes ($E_g$ modes) read:
\begin{equation}
F_{R,i}(t)=\frac{v \epsilon_0}{2} p^2 E_0(t) 
\left[\delta_{i,1} \cos(2\Phi) - \delta_{i,2}\sin(2\Phi)\right]
\int_{-\infty}^t c(t-t^\prime) E_0(t^\prime) dt^\prime ,
\end{equation}
where $a(t)=\partial \chi_{11}(t)/\partial Q_{R,0}$ and 
$c(t)=\partial \chi_{11}(t)/\partial Q_{R,1}=-\partial \chi_{12}(t)/\partial Q_{R,2}$. 
\begin{figure}[!htbp]
\begin{center}
\vskip1.0truecm
\includegraphics[angle=0, scale=0.4]
{./FIG-9-PRB.eps}
\end{center}
%    \centering
%    \includegraphics[width=0.98\textwidth]{Fig1-reply.eps}
\caption{(a) and (b): Real (thick black line) and imaginary (dashed red line)
part of the Raman tensor
$\hat{a}(\omega)$ and $\hat{c}(\omega)$ (in pm$^{-1}$) for 
the A$_{1g}$ and E$_g$ modes as a function
of $\hbar \omega$ (in eV).
(c) and (d): Calculated time evolution of the A$_{1g}$ and E$_g$ mode 
coordinates
following the arrival of the experimental THz pulse polarized along the x-axis 
	($\Phi=0$)
including (thick black curve) and excluding
(thin blue curve) the electronic Raman force. 
Note that the blue curves
correspond to the solid lines in Fig. \ref{dynamics-fig}(b-c). 
%and that the dynamics
%does not depend on the electric field polarization. 
The grey shaded areas
represent the envelope of the THz pulse.
(e) and (f): Forces (in meV.\AA$^{-1}$) acting on the A$_{1g}$ 
and E$_g$ modes arising the lattice anharmonicity
%the anharmonic
%coupling between the E$_u^1$ mode and the A$_{1g}$ mode (Ionic Raman process),
denoted as $F_{R,0}^{anh}$ and $F_{R,1}^{anh}$
(see the thin blue curves), together with the forces arising from a two photon
absorption process (Electronic Raman process), denoted as $F_{R,0}$ and
$F_{R,1}$ (see the thick black curves),
as a function of time $t$ (in ps). The grey shaded area represents
the square of the experimental THz pulse.
}
\label{Electronic_Raman_fig}
\end{figure}
%Here, we focus on the force acting on the A$_{1g}$ mode given by Eq. \ref{F_R0_Raman}.
From a practical point of view, the Fourier transform of $a(t)$ and $c(t)$, 
respectively denoted as 
$\hat{a}(\omega)$ and $\hat{c}(\omega)$, 
%is computed using a finite difference method where 
%$\hat{\chi}_{1,1}(\omega)$ is computed for $Q_{R,0}=\pm 0.8$ pm.
are computed using a finite difference scheme.
%where $Q_{R,i}=\pm 0.8$ pm.
Both the real and imaginary part of the Raman tensors $\hat{a}(\omega)$
and $\hat{c}(\omega)$
are respectively displayed in Fig. \ref{Electronic_Raman_fig}(a)
and Fig. \ref{Electronic_Raman_fig}(b). As can be inferred
from these curves, the real part (imaginary part) of the Raman tensor is constant (negligible)
in the frequency range where the spectral weight of the Fourier transform of the experimental
THz pulse is non negligible, namely between $0$ and $1.5$ THz as
illustrated in Fig. \ref{fig2}(b).
Thus, $a(t-t^\prime)=\hat{a}(0) \delta(t-t^\prime)$ and 
$c(t-t^\prime)=\hat{c}(0) \delta(t-t^\prime)$. Consequently, the
force acting on the A$_{1g}$ mode reads:
\begin{equation}\label{F_R0_Raman_simplified}
F_{R,0}(t)=\frac{v \epsilon_0}{2}  p^2 \hat{a}(0) E_0(t)^2, 
\end{equation}
while the forces acting on the E$_g$ modes read:
\begin{equation}\label{F_Ri_Raman_simplified}
F_{R,i}(t)=\frac{v \epsilon_0}{2} p^2 \hat{c}(0) E_0(t)^2 
\left[\delta_{i,1} \cos(2\Phi) - \delta_{i,2}\sin(2\Phi)\right].
\end{equation}
As shown in Fig. \ref{Electronic_Raman_fig}(e-f), 
the electronic Raman force $F_{R,0}$ ($F_{R,1}$)
is approximately $8.5$ ($5$) times smaller than the ionic Raman force
%$F_{R,0}^{anh}$ 
extracted from our simulations (see Fig. \ref{dynamics-fig}(b-c)).
Figure \ref{Electronic_Raman_fig}(c) demonstrates that the 
inclusion of the electronic Raman force has a negligible effect on the 
A$_{1g}$ mode dynamics, thereby justifying its omission in the simulations
presented in section \ref{Dynamics_modes_part}. 
As shown in Fig. \ref{Electronic_Raman_fig}(d), 
a similar conclusion applies to the E$_g$ modes 
since the oscillation amplitude is only slightly enhanced, 
despite the fortuitous resonance between the E$_g$ mode frequency 
and twice the central frequency of the experimental THz pulse.

%It is important to state that the THz pulse can redistribute 
%the free electrons which can in turn exert a non-equilibrium force on both
%A$_{1g}$ mode and E$_g$ modes. 
From a theoretical point of view, taking into account the bulk charge carriers
originating from intrinsic defects such as anion vacancies or antisite defects
remains elusive. The charge carrier concentration in our Bi$_2$Te$_3$ nanoscale
thin film is unknown but different experimental studies 
show that a p-type or n-type charge
carrier concentration can range from $3\times 10^{17}$ cm$^{-3}$ 
($~5\times 10^{-5}$ carriers per unit cell) to $5\times 10^{19}$
cm$^{-3}$ ($~8.45 \times 10^{-3}$ carriers per unit cell) 
depending on the growth conditions.  It is worth highlighting that 
the number of carriers per unit cell is relatively small and that the
redistribution of these carriers induced by the THz pulse might change
the forces acting on the atoms and compete with the previously discussed forces.
Although, we cannot completely rule out such a scenario, we assert that this
process is negligible, at least for the A$_{1g}$ mode. This is supported
by the fact that the oscillatory
part of the transmittivity, probed at $400$ nm, is accurately
reproduced in our calculations (see Fig. \ref{optics-fig}(c)), underscoring
the importance of modeling the detection process to draw reliable conclusions.

%The dynamics of the A$_{1g}$ mode displayed in Fig. \ref{Electronic_Raman_fig}(c)
%demonstrates that $Q_{R,0}$ is only slightly affected
%by the Raman force of electronic origin, thus justifying our choice to neglect
%it in our calculation.

\section{Analytical results for the IR active modes}\label{analytical_part}
The simplified equation of motion for one of the 
IR active mode ($E_u^1$ mode) reads:
\begin{equation}\label{general_eq_IR}
\ddot{Q} + 2 \gamma_{IR} \dot{Q} + \Omega_{IR}^2 Q = \frac{F_0}{M} 
\sin\left[\omega_0 t + \Phi_0 \right] exp\left[-t^2/\sigma^2\right],
\end{equation}
where $Q=Q_{IR,1}$ or $Q_{IR,2}$ and $F_0=e Z_\perp p E_0$. The Green's function
for the second order homogeneous linear equation reads:
\begin{equation}
G(t)=\theta(t) \frac{\sin(\tilde{\omega} t)}{\tilde{\omega}} \exp\left[-\gamma_{IR} t\right],
\end{equation}
where $\theta(t)$ is the heaviside function and 
$\tilde{\omega}=\sqrt{\Omega_{IR}^2-\gamma_{IR}^2 }$. Thus, the solution
of Eq. \ref{general_eq_IR} reads:
\begin{equation}
Q(t)=\frac{F_0}{M} \int_{-\infty}^t dt^\prime~ G(t-t^\prime) 
\sin\left[\omega_0 t^\prime + \Phi_0 \right] exp\left[-{t^\prime}^2/\sigma^2\right].
\end{equation} 
After straightforward calculations, we get:
\begin{equation}\label{Eq_for_Q}
Q(t)=\frac{F_0}{2 M \tilde{\omega}} \textrm{Re}\left[
\exp\left[i(\tilde{\omega} t - \Phi)\right] A_{+}(t)
-\exp\left[i(\tilde{\omega} t + \Phi)\right] A_{-}(t)
\right],
\end{equation} 
where $A_{\pm}(t)$ is defined by:
%\begin{equation}\label{def_A_pm}
\begin{eqnarray}\label{def_A_pm}
A_{\pm}(t)=\frac{\sigma \sqrt{\pi}}{2}
\exp\left[\frac{\sigma^2 \gamma_{IR}^2}{4}   \right]
\exp\left[-\frac{\sigma^2}{4} \Delta \omega_{\pm}^2   \right]
\exp\left[-i\frac{\sigma^2 \gamma_{IR}}{2} \Delta\omega_{\pm}   \right] \nonumber \\
\times\left\{1 + \textrm{erf}\left[\frac{t}{\sigma} + 
\frac{\sigma}{2}(-\gamma_{IR} + i \Delta\omega_{\pm})    \right]   \right\},
\end{eqnarray}
%\end{equation}
where $\Delta\omega_{\pm}=\tilde{\omega} \pm \omega_0$ and $z \mapsto \textrm{erf}(z)$ 
is the complex error function defined by:
\begin{equation}
\textrm{erf}(z)= \frac{2}{\sqrt{\pi}} \int_0^z dx \exp\left[-x^2\right].
\end{equation}
As $\sigma(\tilde{\omega} + \omega_0)/2 \gg 1$, we conclude that $A_+(t) \simeq 0$.
Thus, Eq. \ref{Eq_for_Q} leads to:
\begin{equation}\label{Eq_for_Q_2}
Q(t)=-\frac{F_0}{2 M \tilde{\omega}} \textrm{Re}\left[
\exp\left[i(\tilde{\omega} t + \Phi)\right] A_{-}(t) \right].
\end{equation} 
As $\textrm{erf}\left[t/\sigma + \sigma(-\gamma_{IR} + i \Delta\omega_{-})/2    \right] 
\to 1$ when $t \to \infty$, 
the long time behaviour of the phonon coordinate $Q$ is given by:
\begin{equation}\label{long_time_behaviour_QIR}
Q^{long}(t)= -\frac{F_0 \sigma \sqrt{\pi}}{2 M \tilde{\omega}}
\exp\left[-\gamma_{IR} t\right] 
\exp\left[-\frac{\sigma^2}{4}\Delta\omega_{-}^2   \right]
\exp\left[\frac{\sigma^2\gamma_{IR}^2}{4}\right]
\cos\left[\tilde{\omega} t + \Psi\right],
\end{equation}
where $\Psi=\Phi_0 -\gamma_{IR} \sigma^2 \Delta\omega_{-}/2$. As stated 
in section \ref{Dynamics_modes_part}, 
the oscillation amplitude is very small for $t>2\sigma$ because
the THz pulse is non resonant with the IR mode ($\Delta\omega_{-}\gg 2/\sigma$).

\section{Dynamics of the modes for an hypothetical THz pulse}\label{hypothetical_pulse_part}
The dynamics of the modes shown in Fig \ref{dynamics-fig} have been 
obtained for the parameters describing the 
experimental THz pulse (See section \ref{expt-part}). Here,
we change the central frequency $\omega_0/2\pi$ of the THz pulse while 
keeping the other parameters unchanged and explore how the dynamics of the
modes are affected.  
\begin{figure}[!htbp]
\vskip0.8truecm
\includegraphics[angle=0, scale=0.38]
{./FIG-10-PRB.eps}
%    \centering 
%    \includegraphics[width=0.98\textwidth]{Figure_Fit_SM.eps}
\caption{
Calculated time evolution of the E$_{u}^1$ (a),
A$_{1g}^1$ (b) and E$_{g}^1$ (c) mode coordinates following
the arrival of a THz pulse polarized along the x-axis ($\Phi=0$) with an amplitude
$\sim$ 193 kV.cm$^{-1}$ inside the Bi$_2$Te$_3$ nanofilm ($E_0 \sim 340$ kV.cm$^{-1}$) 
with $\Phi_0=0.117 \pi$ and $\omega_0/2\pi=0.93$ THz.
The grey shaded areas represent the envelope of the THz pulse.
The normalized Fourier Transforms (FT) of the phonon coordinates $Q$, denoted as $\hat{Q}$,
are shown in panels (d), (e) and (f).
The FT of the THz electric field is shown as a dashed line in panel (d) and the long time
behaviour of the E$_{u}^1$ mode (See Eq. \ref{long_time_behaviour_QIR}) 
is displayed as a dotted line in panel (a).
}
\label{Dynamics_0.93THz_fig}
\end{figure}
We first consider the case where $\omega_0/2\pi$ is tuned to half the frequency 
of the $A_{1g}^1$ mode. The dynamics of the $E_u$, $A_{1g}$ and $E_g$ modes following
the arrival of the THz pulse are displayed in Fig. \ref{Dynamics_0.93THz_fig}. As compared
to the curve shown in Fig. \ref{dynamics-fig}(a), the IR active mode
is more efficiently driven when the central frequency of the THz pulse is closer 
to the frequency of the $E_u$ mode. The oscillations of the $E_u$ mode after the end 
of the THz pulse are especially noticeable and well described by Eq. \ref{long_time_behaviour_QIR}.
Indeed, the analytical result corresponding to the long time behavior of the IR active
mode (See the dotted line) matches the computed dynamics (See the straight line) 
when $t>2\sigma \sim 1.6$ ps. Concomitantly, the Fourier Transform (FT) of the IR coordinate
exhibits a peak around 1.5 THz (See Fig. \ref{Dynamics_0.93THz_fig}(d)) 
while this peak was hardly seen for the experimental THz pulse 
(See Fig \ref{dynamics-fig}(d)). 
The dynamics of the $A_{1g}$ mode shown in Fig. \ref{Dynamics_0.93THz_fig}(b) is very 
similar to the dynamics shown in Fig. \ref{dynamics-fig}(b). 
The only difference 
is that $Q_{R,0}$ is roughly scaled by a factor $\sim 2.4$. Hence,
$\left|Q_{R,0}\right|_{max}\sim 0.14$ pm while $\left|Q_{R,0}\right|_{max}\sim 0.06$ pm
for the experimental THz pulse. Such a difference arises not only from the fact that the IR mode
is more efficiently driven but also from the fact that the double of the frequency of the
THz pulse matches the frequency of the $A_{1g}$ mode. 
%As a matter of fact, the energy 
%transfered to the $A_{1g}$ mode, denoted as $\Delta E$, is given by:
%\begin{equation}
%\Delta E = \frac{1}{2M} \left|\widehat{F_{R,0}^{anh}}(\Omega_{R,0})\right|^2,
%\end{equation}
%provided that $\gamma_{R,0} \to 0$ (No damping of the phonon). As 
%$F_{R,0}^{anh}=M g_0 Q_{IR,1}^2$, we can conclude that $\Delta E$ is maximum
%when the spectral weight of $\widehat{Q_{IR,1}^2}(\omega)$ (FT of $Q_{IR,1}^2(t)$) 
%is centered on $\Omega_{R,0}$.
Eventually, the dynamics of the $E_{1g}^1$ mode shown in Fig. \ref{Dynamics_0.93THz_fig}(c)
is different from the dynamics shown in Fig. \ref{dynamics-fig}(c). Indeed, 
the displacement of the $Q_{R,1}$ coordinate during the THz pulse, arising from phononic
rectification, is slightly enhanced and 
the amplitude
of the oscillations after the end of the THz pulse is five time smaller in accordance 
with the fact that the double of the central frequency of the THz pulse 
does not coincide anymore with the frequency of the $E_{1g}^1$ mode.
\begin{figure}[!htbp]
\vskip0.8truecm
\includegraphics[angle=0, scale=0.38]
{./FIG-11-PRB.eps}
%    \centering 
%    \includegraphics[width=0.98\textwidth]{Figure_Fit_SM.eps}
\caption{
Calculated time evolution of the E$_{u}^1$ (a),
A$_{1g}^1$ (b) and E$_{g}^1$ (c) mode coordinates following
the arrival of a THz pulse polarized along the x-axis ($\Phi=0$) with an amplitude
$\sim$ 193 kV.cm$^{-1}$ inside the Bi$_2$Te$_3$ nanofilm ($E_0 \sim 340$ kV.cm$^{-1}$) 
with $\Phi_0=0.117 \pi$ and $\omega_0/2\pi=1.5$ THz.
The grey shaded areas represent the envelope of the THz pulse.
The normalized Fourier Transforms (FT) of the phonon coordinates $Q$, denoted as $\hat{Q}$,
are shown in panels (d), (e) and (f).
The FT of the THz electric field is shown as a dashed line in panel (d) and the long time
behaviour of the E$_{u}^1$ mode (See Eq. \ref{long_time_behaviour_QIR}) 
is displayed as a dotted line in panel (a).
}
\label{Dynamics_1.5THz_fig}
\end{figure}

Now we consider the case where the THz pulse is resonant with the $E_u^1$ mode
($\omega_0/2\pi=1.5$ THz). The dynamics of the $E_u^1$, $A_{1g}$ and $E_g^1$ modes
are displayed in Fig. \ref{Dynamics_1.5THz_fig}. As expected, the IR active mode 
is strongly driven by the THz pulse and the amplitude of oscillations is four times
larger than for the experimental THz pulse. Again, the long time behaviour of the
$E_u^1$ mode is well described by Eq. \ref{long_time_behaviour_QIR} assessing the
validity of our analytical approach. Interestingly, the $Q_{R,0}$ and $Q_{R,1}$
coordinates, respectively displayed in Fig. \ref{Dynamics_1.5THz_fig}(b) and 
Fig. \ref{Dynamics_1.5THz_fig}(c), are not oscillating anymore around their
equilibrium position up to 5 ps. Thus, the crystal symmetry is transiently 
lowered, through a nonlinear phononic rectification mechanism, to 2/m 
(This group has $4$ symmetry operations, compared to the $12$ 
symmetry operations of the D$_{3d}$ point group) when
the $Q_{R,1}$ mode is coherently driven by a THz pulse polarized along the $x-$axis 
($\Phi=0$). It's worth pointing out that the $Q_{R,1}$ mode is not driven anymore
when the polarization of the electric field is rotated from 45$^\circ$ ($\Phi=\pi/4$).
Instead, the $Q_{R,2}$ mode is coherently driven and the crystal symmetry is transiently
lowered to $\overline{1}$ (This group has $2$ symmetry operations). 
In conclusion, the polarization of a THz pulse 
resonant with the $E_u^1$ mode might offer the unique
opportunity to steer the dynamics of atoms along the $E_g^1$ mode and 
to distort the crystal according to a given symmetry.

\section{CONCLUSION}\label{conclusion_part}
Our first-principles calculations unravel the mechanisms behind the generation
of the $A_{1g}^1$ phonon in THz excited Bi$_2$Te$_3$ nanofilms. Indeed, 
the transient electric field non-resonantly drives the $E_u^1$ mode which in 
turn is coupled to both $A_{1g}^1$ and $E_g^1$ modes through phonon-phonon 
interactions arising from cubic terms in the total energy expansion. Thus,
the generation of the $A_{1g}^1$ mode can not be ascribed to quartic terms,
as speculated for Bi$_2$Se$_3$\cite{Melnikov_2018}, a parent coumpound of Bi$_2$Te$_3$.
Regarding the detection process, we demonstrate that the optical properties
of the nanofilm are modulated by the $A_{1g}$ mode and show that we can achieve
a quantitative description of the oscillatory part of the transient transmittance
using the characteristics of both pump and probe pulses and, remarkably without 
introducing any adjustable parameters in our calculations. Furthermore, the atomic
displacements can be inferred from {\textit{ab initio}} calculations without resorting to
THz-pump and X-ray probe experiments as done in the study of THz-driven upconversion
in SrTiO$_3$\cite{Kozina_2019}. Finally, we predict that a THz pump with
a frequency matching the frequency of the $E_u^1$ mode can transiently lower 
the symmetry of the system.

\begin{acknowledgments}
We acknowledge GENCI-CINES (project 095096) for high performance
computing resources and the
funding from the French National Research Agency (ANR) 
through the EPHONO project (Grant No. ANR-22-CE30-0007). 
We also thank Dr. K. Balin
for providing high quality Bi$_2$Te$_3$ thin films grown by MBE 
at the Physics Institute of Silesia University.
\end{acknowledgments}

\end{document}